\documentclass[twocolumn,showpacs,prd,preprintnumbers,superscriptaddress]{revtex4}

\usepackage{graphicx}
\usepackage{epsfig}
\usepackage{xspace}
\usepackage{dcolumn}
\usepackage{multirow}
\usepackage{longtable}

\newcommand{\superk}      {Super-Kamiokande\xspace}       

\newcommand{\evis}        {$E_{vis}$}
\newcommand{\nue}         {$\nu_{e}$\xspace}
\newcommand{\numu}        {$\nu_{\mu}$\xspace}

\newcommand{\tonethree}   {$\theta_{13}$\xspace}
\newcommand{\ttwothree}   {$\theta_{23}$\xspace}

\newcommand{\SK}          {Super-K\xspace}
\newcommand{\upmu}        {UP$\mu$\xspace}

\newcolumntype{d}[1]{D{.}{\cdot}{#1}}
\newcolumntype{.}{D{.}{.}{-1}}
\newcolumntype{,}{D{,}{,}{2}}
\begin{document}

\title{ Atmospheric neutrino oscillation analysis with
        sub-leading effects in Super-Kamiokande I, II, and III  }

\newcommand{\AFFicrr}{\affiliation{Kamioka Observatory, Institute for Cosmic Ray Research, University of Tokyo, Kamioka, Gifu 506-1205, Japan}}
\newcommand{\AFFkashiwa}{\affiliation{Research Center for Cosmic Neutrinos, Institute for Cosmic Ray Research, University of Tokyo, Kashiwa, Chiba 277-8582, Japan}}
\newcommand{\AFFipmu}{\affiliation{Institute for the Physics and
Mathematics of the Universe, University of Tokyo, Kashiwa, Chiba
277-8582, Japan}}
\newcommand{\AFFbu}{\affiliation{Department of Physics, Boston University, Boston, MA 02215, USA}}
\newcommand{\AFFbnl}{\affiliation{Physics Department, Brookhaven National Laboratory, Upton, NY 11973, USA}}
\newcommand{\AFFucd}{\affiliation{Department of Physics, University of California, Davis, Davis, CA 95616, USA}}
\newcommand{\AFFuci}{\affiliation{Department of Physics and Astronomy, University of California, Irvine, Irvine, CA 92697-4575, USA }}
\newcommand{\AFFcsu}{\affiliation{Department of Physics, California State University, Dominguez Hills, Carson, CA 90747, USA}}
\newcommand{\AFFcnm}{\affiliation{Department of Physics, Chonnam National University, Kwangju 500-757, Korea}}
\newcommand{\AFFduke}{\affiliation{Department of Physics, Duke University, Durham NC 27708, USA}}
\newcommand{\AFFgmu}{\affiliation{Department of Physics, George Mason University, Fairfax, VA 22030, USA }}
\newcommand{\AFFgifu}{\affiliation{Department of Physics, Gifu University, Gifu, Gifu 501-1193, Japan}}
\newcommand{\AFFuh}{\affiliation{Department of Physics and Astronomy, University of Hawaii, Honolulu, HI 96822, USA}}
\newcommand{\AFFkanagawa}{\affiliation{Physics Division, Department of Engineering, Kanagawa University, Kanagawa, Yokohama 221-8686, Japan}}
\newcommand{\AFFkek}{\affiliation{High Energy Accelerator Research Organization (KEK), Tsukuba, Ibaraki 305-0801, Japan }}
\newcommand{\AFFkobe}{\affiliation{Department of Physics, Kobe University, Kobe, Hyogo 657-8501, Japan}}
\newcommand{\AFFkyoto}{\affiliation{Department of Physics, Kyoto University, Kyoto, Kyoto 606-8502, Japan}}
\newcommand{\AFFumd}{\affiliation{Department of Physics, University of Maryland, College Park, MD 20742, USA }}
\newcommand{\AFFmit}{\affiliation{Department of Physics, Massachusetts Institute of Technology, Cambridge, MA 02139, USA}}
\newcommand{\AFFmiyagi}{\affiliation{Department of Physics, Miyagi University of Education, Sendai, Miyagi 980-0845, Japan}}
\newcommand{\AFFnagoya}{\affiliation{Solar Terrestrial Environment
Laboratory, Nagoya University, Nagoya, Aichi 464-8602, Japan}}
\newcommand{\AFFsuny}{\affiliation{Department of Physics and Astronomy, State University of New York, Stony Brook, NY 11794-3800, USA}}
\newcommand{\AFFniigata}{\affiliation{Department of Physics, Niigata University, Niigata, Niigata 950-2181, Japan }}
\newcommand{\AFFokayama}{\affiliation{Department of Physics, Okayama University, Okayama, Okayama 700-8530, Japan }}
\newcommand{\AFFosaka}{\affiliation{Department of Physics, Osaka University, Toyonaka, Osaka 560-0043, Japan}}
\newcommand{\AFFseoul}{\affiliation{Department of Physics, Seoul National University, Seoul 151-742, Korea}}
\newcommand{\AFFshizuokasc}{\affiliation{Department of Informatics in
Social Welfare, Shizuoka University of Welfare, Yaizu, Shizuoka, 425-8611, Japan}}
\newcommand{\AFFshizuoka}{\affiliation{Department of Systems Engineering, Shizuoka University, Hamamatsu, Shizuoka 432-8561, Japan}}
\newcommand{\AFFskk}{\affiliation{Department of Physics, Sungkyunkwan University, Suwon 440-746, Korea}}
\newcommand{\AFFtohoku}{\affiliation{Research Center for Neutrino Science, Tohoku University, Sendai, Miyagi 980-8578, Japan}}
\newcommand{\AFFtokyo}{\affiliation{The University of Tokyo, Bunkyo, Tokyo 113-0033, Japan }}
\newcommand{\AFFtokai}{\affiliation{Department of Physics, Tokai University, Hiratsuka, Kanagawa 259-1292, Japan}}
\newcommand{\AFFtit}{\affiliation{Department of Physics, Tokyo Institute
for Technology, Meguro, Tokyo 152-8551, Japan }}
\newcommand{\AFFtsinghua}{\affiliation{Department of Engineering Physics, Tsinghua University, Beijing, 100084, China}}
\newcommand{\AFFwarsaw}{\affiliation{Institute of Experimental Physics, Warsaw University, 00-681 Warsaw, Poland }}
\newcommand{\AFFuw}{\affiliation{Department of Physics, University of Washington, Seattle, WA 98195-1560, USA}}

\AFFicrr
\AFFkashiwa
\AFFipmu
\AFFbu
\AFFbnl
\AFFuci
\AFFcsu
\AFFcnm
\AFFduke
\AFFgifu
\AFFuh
\AFFkanagawa
\AFFkek
\AFFkobe
\AFFkyoto
\AFFmiyagi
\AFFnagoya
\AFFsuny
\AFFniigata
\AFFokayama
\AFFosaka
\AFFseoul
\AFFshizuoka
\AFFshizuokasc
\AFFskk
\AFFtokai
\AFFtokyo
\AFFtsinghua
\AFFwarsaw
\AFFuw
%
\author{R.~Wendell}
\AFFduke

\author{C.~Ishihara}
\AFFkashiwa

\author{K.~Abe}
\AFFicrr
\author{Y.~Hayato}
\AFFicrr
\AFFipmu
\author{T.~Iida}
\author{M.~Ikeda}
\author{K.~Iyogi} 
\author{J.~Kameda}
\author{K.~Kobayashi}
\author{Y.~Koshio}
\author{Y.~Kozuma} 
\author{M.~Miura} 
\AFFicrr
% ...

\collaboration{The Super-Kamiokande Collaboration}
\noaffiliation

\date{\today}

\begin{abstract}

We present a search for non-zero \tonethree and deviations 
of $\mbox{sin}^{2} \theta_{23}$ from 0.5  in the oscillations of atmospheric 
neutrino data from \superk -I,  -II, and -III. 
No distortions of the neutrino flux consistent with non-zero \tonethree are found
and both neutrino mass hierarchy hypotheses are in agreement with the data. 
The data are best fit at $\Delta m^{2} = 2.1 \times 10^{-3}~\mbox{eV}^{2}$,
$\mbox{sin}^{2}\,\theta_{13} = 0.0 $, and $\mbox{sin}^{2} \theta_{23} =0.5.$
In the normal (inverted) hierarchy \tonethree and $\Delta m^{2}$ 
are constrained at the one-dimensional 90\% C.L. to  $\mbox{sin}^{2} \theta_{13} < 0.04 \xspace ( 0.09 )$ 
and $1.9 (1.7) \times 10^{-3} < \Delta m^{2} < 2.6 (2.7) \times 10^{-3} \mbox{eV}^{2}.$
The atmospheric mixing angle is within $ 0.407 \le \mbox{sin}^{2} \theta_{23} \le 0.583 $ at 90\% C.L.

\end{abstract}


\maketitle

\section{Introduction}

Despite experimental measurements of 
solar~\cite{Homestake98, SAGE09, GALLEX99, GNO05, Hosaka:2005um, Cravens08,SNO05,SNOLETA}, reactor~\cite{KamLAND08},  
atmospheric~\cite{Hosaka:2006zd,ashie:2005ik}, and accelerator~\cite{K2K06:prd,Minos08} neutrinos
constraining their flavor oscillations, the nature of the neutrino mass hierarchy and whether or not 
\tonethree is zero remain open questions. The latter is the last unknown mixing angle 
and is currently the subject of a research program including beam and reactor-based 
experiments~\cite{Itow:2001ee, Ardellier:2006dc,Dayabay:2007doe,NOvA:2005,RENO:2008,Komatsu:2002sz}.
At present, experiments have placed upper limits on the value of \tonethree~\cite{K2K06:q13,PaloVerde01,Hosaka:2006zd,Minos09} 
with the most stringent limit set by the Chooz~\cite{Chooz03} experiment.
However, a non-zero value may manifest itself and be observable in the event rate of multi-GeV 
electron neutrinos passing through the Earth, and to a lesser extent, in similarly energetic upward-going muon samples.
Though atmospheric neutrino data are well fit to 
pure $\nu_{\mu} \leftrightarrow \nu_{\tau}$ oscillations with ``maximal atmospheric mixing''~\cite{ashie:2005ik}
($\theta_{23} = \pi/4),$
$ \nu_{\mu} \leftrightarrow \nu_{e}$ transitions driven by solar oscillation parameters
appear at sub-GeV energies when the atmospheric mixing deviates from this value.
The questions of whether or not \ttwothree is exactly $\pi / 4,$ the nature of \tonethree, and 
the sign of the neutrino mass hierarchy all contribute to an eight-fold 
degeneracy~\cite{Barger02} of oscillation parameter solutions when considering
CP-violation in neutrinos. For future experimental searches of CP-violation, answers to these
questions are essential.

In this paper two analyses are presented searching for evidence of sub-leading (second order) 
oscillation effects which address these questions and appear as changes in 
the \nue and \numu fluxes of the atmospheric neutrino samples at \superk (\SK, SK).
The first is an improved extension of a three flavor oscillation analysis 
using the first phase of the experiment (SK-I)~\cite{Hosaka:2006zd}. 
An updated analysis using the first, second (SK-II), and third (SK-III) phases is presented here. 
The data are then used in the second analysis to test whether 
$\theta_{23}$ deviates from  $\pi /4$.

The paper is organized as follows. In section~\ref{sec:oscillations} we describe 
the oscillation framework used in the analyses. Section~\ref{sec:datasample}
discusses the data sample including additional sample selections designed to
improve the sensitivity of each analysis. The methods and results 
of both are then presented in section~\ref{sec:analysis} and concluding 
remarks are found in section~\ref{sec:conclusion}.

\section{Sub-dominant effects in atmospheric neutrino oscillations}
\label{sec:oscillations}

  Neutrino oscillations in three flavors are described by six parameters: two mass squared differences,
$\Delta m^{2}_{12},$ $\Delta m^{2}_{13},$ 
where  $\Delta m^{2}_{ij}= m^{2}_{j} - m^{2}_{i},$ a CP violating parameter $\delta_{\mbox{cp}}$, 
and three mixing angles $\theta_{ij}, (i < j)$. Each mixing angle parameterizes a rotation, $U_{ij}$,
between mass states inside of the three-dimensional oscillation space. 
The correspondence between neutrino mass eigenstates and their flavor eigenstates is then:

%
\begin{equation}
\vert \nu_{\alpha} \rangle = \sum_{i}^{3} U^{*}_{\alpha, i} \vert \nu_i \rangle, \nonumber
\label{eq:mass2flavor}
\end{equation}

\noindent where $U$ is the $3 \times 3$ PMNS matrix~\cite{Pontecorvo:1967fh,Maki:1962mu} defined by $U_{23}U_{13}U_{12},$
\begin{eqnarray}
U  =
\left(
   \begin{array}{ccc}
   1 & 0 & 0\\
   0 & c_{23} & s_{23} \\
   0 & -s_{23} & c_{23}
   \end{array}
\right)
&
\left(
   \begin{array}{ccc}
   c_{13} & 0 & s_{13}e^{-i\delta_{cp}}\\
   0 & 1 & 0 \\
   -s_{13}e^{i\delta_{cp}} & 0 & c_{13}
   \end{array}
\right)
\nonumber
\\
&
\times
\left(
   \begin{array}{ccc}
   c_{12} & s_{12} & 0\\
   -s_{12} & c_{12} & 0 \\
   0 & 0 & 1
   \end{array} 
\right).
\label{eq:factoredU}
\end{eqnarray}
Non-zero mixing angles and non-degenerate mass eigenvalues give rise to standard neutrino oscillations. 
Observations of solar and reactor neutrinos 
are well-described by oscillations governed by the ``1-2'' (solar) parameters while
those of atmospheric and accelerator neutrinos are described by the 
``2-3'' (atmospheric) parameters. 
These measurements have established two oscillation frequencies which differ by a factor of $\sim 30$. 
The third set of parameters has been probed by Chooz, a reactor neutrino disappearance 
experiment sensitive to oscillations at the atmospheric $\Delta m^{2},$ which placed 
a limit on mixing in this channel at $\mbox{sin}^{2} \theta_{13} < 0.04 $ 
for $\Delta m^{2} \sim 2.0\times 10^{-3}~\mbox{eV}^{2}$ at 90\% confidence~\cite{Chooz03}. 
\begin{figure*}[htbp]
  \begin{minipage}{2.5in}
    \includegraphics[width=2.4in,type=pdf,ext=.pdf,read=.pdf]{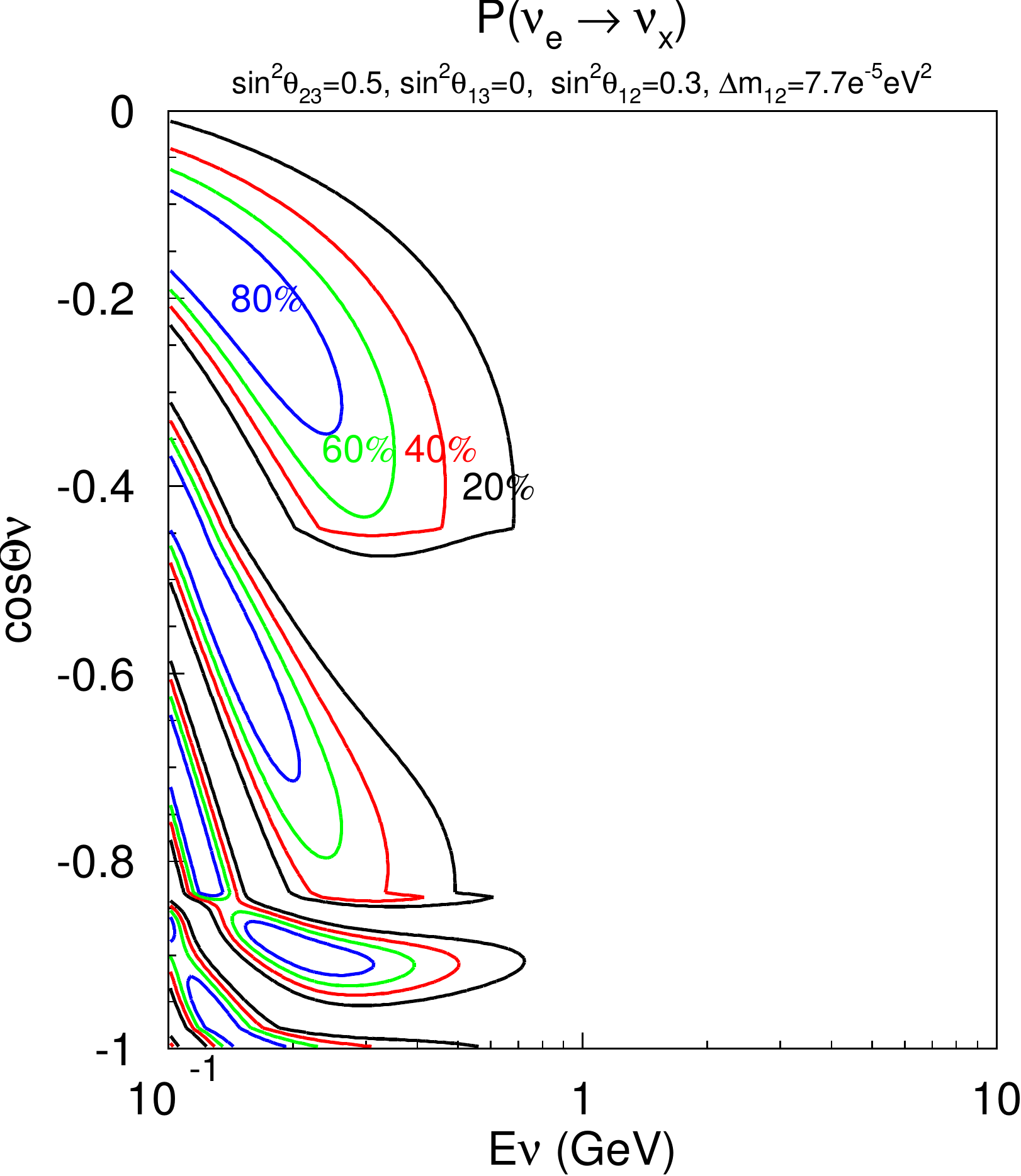}
  \end{minipage}
  \begin{minipage}{2.0in}
    \includegraphics[width=1.8in,type=pdf,ext=.pdf,read=.pdf]{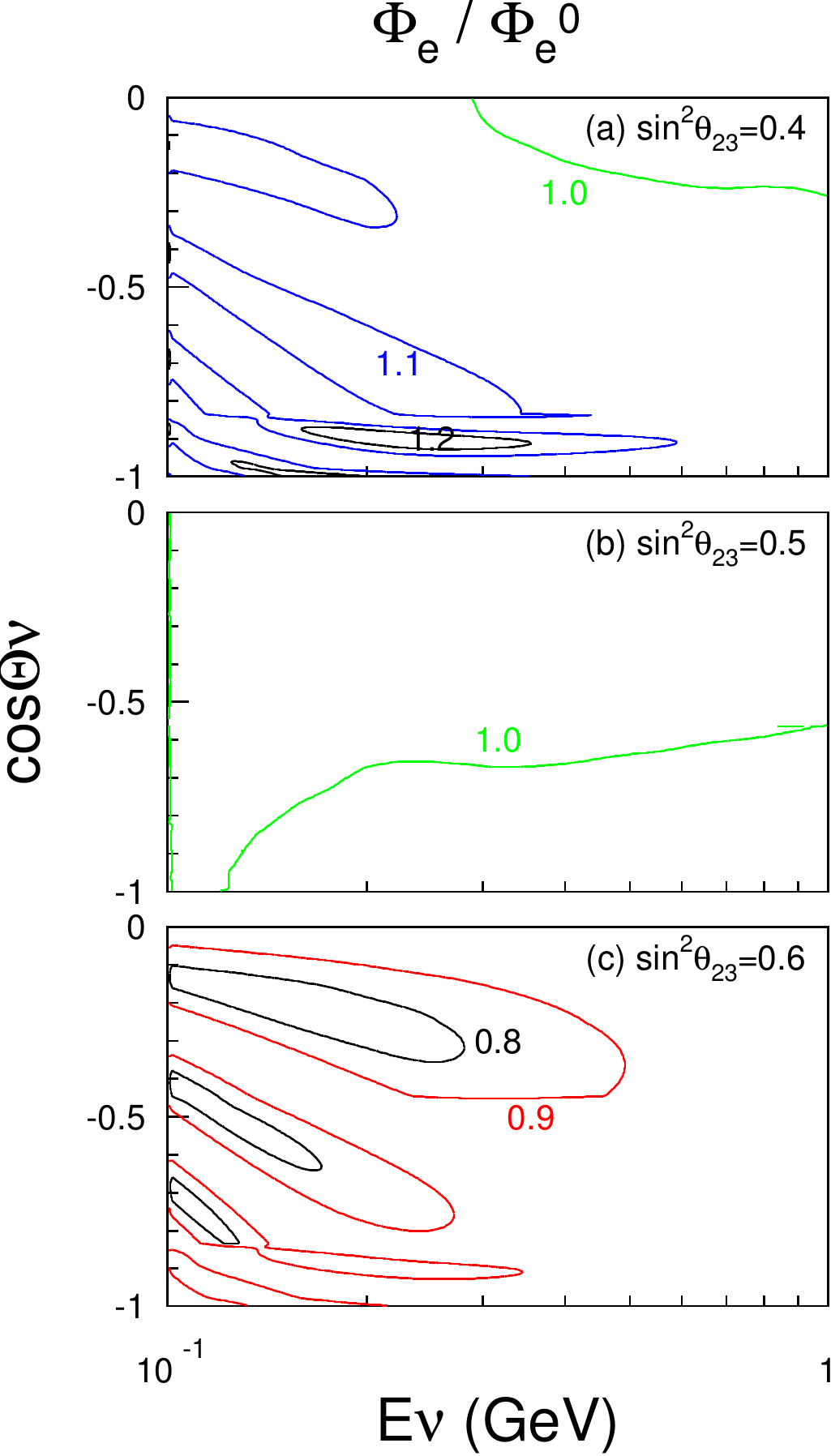}
  \end{minipage}
  \caption{ 
    (color online).
    The left side of the figure shows the calculated $\nu_e$ transition probability $P_{ex}$ for atmospheric neutrinos with 
    an energy $E_\nu$ and neutrino zenith angle, cos$\Theta_{\nu}$, 
    using $\Delta m^{2}_{12} = 7.7 \times 10^{-5} eV^{2} \mbox{sin}^{2}\theta_{12} = 0.3$~\cite{Schwetz08}, $\mbox{sin}^{2}\theta_{23} = 0.5, 
    \mbox{sin}^{2}\theta_{13} = 0.0 \mbox{~and~} \Delta m^{2}_{23} = 2.1 \times 10^{-3} eV^{2}.$ 
    Matter effects within the Earth are taken into account.  
    Negative $\mbox{cos}\Theta_{\nu}$ corresponds to upward-going neutrinos and 0 
    is the horizon. 
    The electron neutrino flux ratio $\Phi_{e}/\Phi_{e}^{0}$ is shown in the right side of the figure.
    An expected excess (deficit) for atmospheric mixing in the first (second) octant 
    is shown in the upper (lower) panel.
    The island shapes are regions of probability driven by the solar oscillation parameters.      
    The center panel shows no significant region of excess or deficit when $\mbox{sin}^{2} \theta_{23} = 0.5.$ 
  }
  \label{fig:prob_nue_x}
\end{figure*}

  For the purposes of studying sub-dominant oscillations in atmospheric neutrinos, it is useful 
to consider oscillation probabilities in two domains: (i) $\theta_{13} \sim 0 $ such that
$U_{13} \sim \mathbf{I}$,  and (ii) $\theta_{13} > 0$, but oscillations driven by the solar parameters
are negligible. The observable effects each domain has on the 
atmospheric neutrino sample can similarly be divided into two energy regimes motivating 
two separate analyses with distinct foci: each analysis has been tailored to 
its regime of interest.  

  In the case of $\theta_{13} \sim 0$, the
neutrino oscillation probabilities in constant density matter may be written~\cite{Peres99}
%
\begin{eqnarray}
P(\nu_{e}   \leftrightarrow \nu_{\mu} )  & = & \mbox{cos}^{2}\theta_{23} P_{ex}  \label{eqn:oscmue}\\
P(\nu_{\mu} \leftrightarrow \nu_{\mu} )  & = & 1 - \mbox{cos}^{4} \theta_{23} P_{ex} \\ \nonumber
                                    &&- \mbox{sin}^{2} 2 \theta_{23} ( 1 - \sqrt{1-P_{ex}} \mbox{cos}{\phi} ) \\ \nonumber
\phi &\sim& \left(\Delta m^2_{31} + s^2_{12} \Delta m^2_{21}\right)\frac{L}{2E_\nu},                        \nonumber
\label{twonusolar} 
\end{eqnarray}
\noindent where $P_{ex}$ is the two neutrino transition probability ($\nu_e \rightarrow \nu_x$)
driven by $\Delta m^{2}_{12}$ and $\theta_{12}$, $L$ is the neutrino pathlength, and $E$ is its energy.
Using these equations the modified atmospheric $\nu$ fluxes at Super-K become: 
%
\begin{eqnarray}
\Phi_{e}   & = & \Phi_{e}^{0}   \left[ 1 + P_{ex} \left( r \mbox{cos}^{2} \theta_{23} - 1 \right) \right] \label{eqn:flx_e_solar} \\
\Phi_{\mu} & = & \Phi_{\mu}^{0} \left[ 1 - \frac{ \mbox{cos}^{2} \theta_{23}}{r} \left( r \mbox{cos}^{2} \theta_{23} - 1 \right) P_{ex} \right] \nonumber \\ 
                  && - \frac{ \Phi_{\mu}^{0} }{2} \mbox{sin}^{2} 2 \theta_{23} 
                        \left( 1 - \sqrt{1-P_{ex}} \mbox{cos}{\phi}\right)  , \nonumber
\label{eqn:fluxsolar}
\end{eqnarray}
\noindent where $\Phi_{\mu}^{0}$ and $\Phi_{e}^{0}$ are the neutrino fluxes in the absence of oscillations and $r$ is their ratio.

  The left panel in Fig.~\ref{fig:prob_nue_x} shows the transition probability, $P_{ex}$, as a function of energy 
and zenith angle, $\Theta_{\nu}$,  for neutrinos traversing the Earth (see below) 
assuming $\Delta m^{2}_{12} = 7.7 \times 10^{-5}~\mbox{eV}^{2}$ and $\mbox{sin}^{2}\theta_{12}$ = 0.30~\cite{Schwetz08}. 
However, the overall effect of $P_{ex}$ on the electron neutrino flux at the detector is modified by the factor
$r \mbox{cos}^{2} \theta_{23} - 1$ as seen in Eq.~(\ref{eqn:flx_e_solar}).
Since the atmospheric neutrino flux ratio is $ r \sim 2$ at low energies, 
there is no change in the $\nu_{e}$ flux if $r \cos^2\theta_{23}=1$ ($\theta_{23}= \pi / 4 $). 
If $\cos^2\theta_{23}$ is greater (less) than 0.5 ($\theta_{23}< (>) \pi/ 4 $) 
there is an expected enhancement (reduction) of the flux.
Therefore it may be possible to determine the octant of $\theta_{23}$ by observing changes in the
flux of the low energy electron-like (\mbox{$e$-like}) samples at SK. Analogous changes to the $\nu_{\mu}$ flux on the other 
hand are suppressed by the leading factor of $1/r$ in the second term of Eq.~(\ref{eqn:flx_e_solar}).  
Further, vacuum oscillations of the low energy $\nu_{\mu}$ flux are already well averaged so the correction appearing in the 
third term is negligible. The expected change in the $\nu_{e}$
flux as a function of energy and  zenith angle for different values of $\theta_{23}$ is 
shown as the right panel of Fig.~\ref{fig:prob_nue_x}. 
\begin{figure*}[htbp]
  \begin{minipage}{6.1in}
    \includegraphics[width=3.0in,type=pdf,ext=.pdf,read=.pdf]{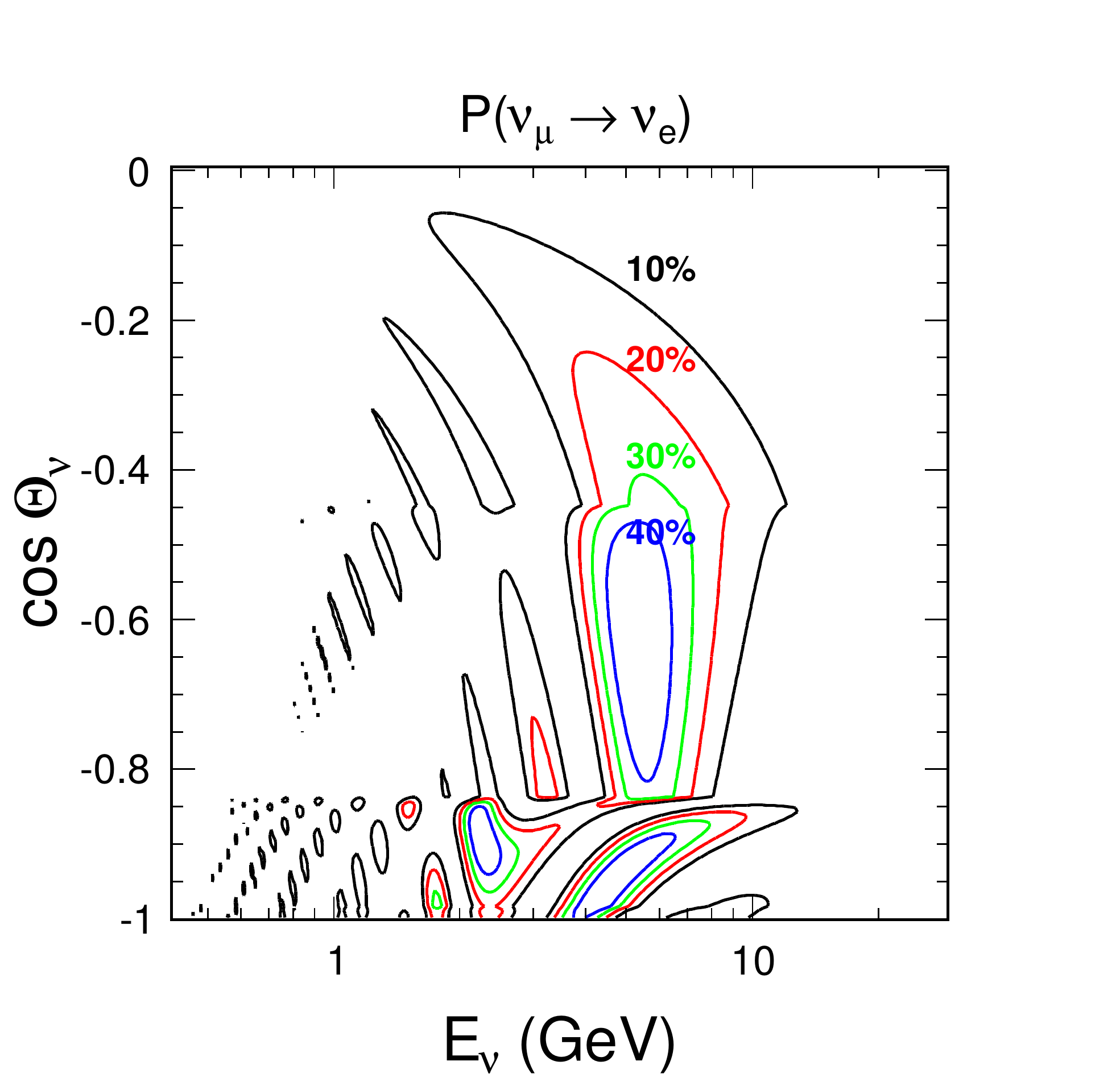}
    \includegraphics[width=3.0in,type=pdf,ext=.pdf,read=.pdf]{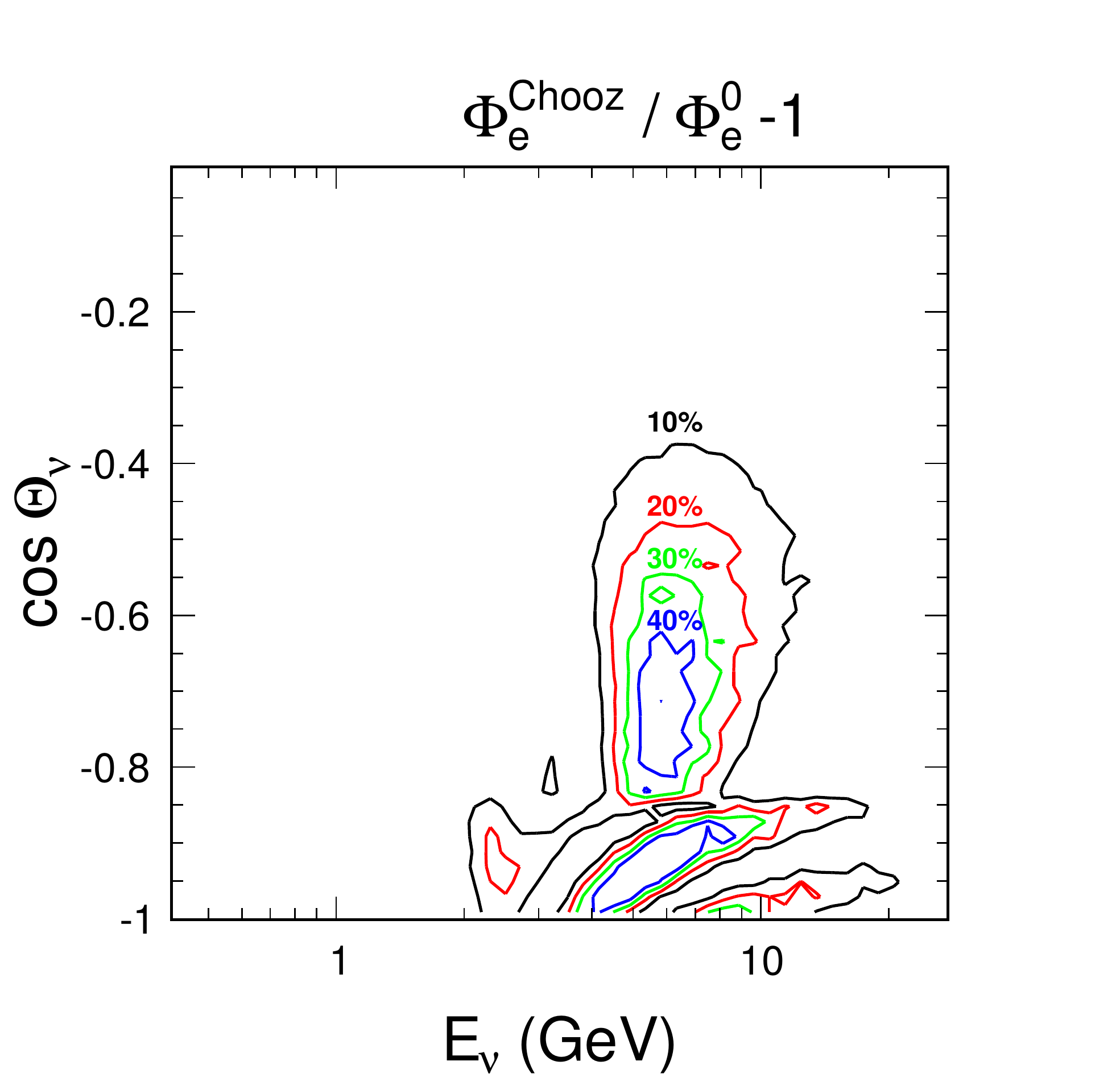}
     \caption{ 
        (color online).
        The three flavor oscillation probability $\nu_{\mu} \leftrightarrow \nu_{e}$ 
        for $\theta_{13}$ at the Chooz limit
        for neutrinos under the normal hierarchy in the one mass scale dominant framework is shown at left. 
        In the right panel the $\nu_{e}$ flux ratio $\Phi_{e}^{Chooz}/\Phi_{e}^{0} - 1$ for oscillations 
        with $\theta_{13}$ at the Chooz limit relative to those at $\theta_{13} = 0.$ 
        Large matter-induced resonances between 2-10 GeV appear for upward going neutrinos 
        traversing the core ($\mbox{cos}\Theta_{\nu} < -0.84$) and mantle regions 
        ($-0.84 < \mbox{cos}\Theta_{\nu} < -0.45 $). Atmospheric mixing is assumed at
        $\Delta m^{2}_{23} = 2.1\times 10^{-3}~\mbox{eV}^{2}$ and $\mbox{sin}^{2} \theta_{23} = 0.5 $.
     }
  \label{fig:3fprob}
  \end{minipage}
\end{figure*}

  When \tonethree is different from zero, the matrix $U_{13}$ is no longer sufficiently close to unity, and the above relations do not hold.
Instead, in the search for non-zero $\theta_{13}$, the oscillation analysis is done using a ``one mass scale dominant'' 
scheme wherein the solar neutrino mass difference is taken to be much smaller than the atmospheric mass difference.
Accordingly, the solar mass difference is neglected and a single mass splitting is adopted, 
$\Delta m^{2} \equiv  m_{3}^{2} - m_{1,2}^{2} $ such that $\Delta m^{2} > 0 ( \Delta m^{2} < 0) $
corresponds to the normal (inverted) mass hierarchy. In vacuum:
%
\begin{eqnarray}
P(\nu_e \rightarrow \nu_e) &=& 1- \mbox{sin}^2 2\theta_{13} \mbox{sin}^2 \left(\frac{1.27 \Delta m^2 L}{E}\right) \nonumber \\
P(\nu_\mu \leftrightarrow \nu_e) &=& \mbox{sin}^2 \theta_{23} \mbox{sin}^2 2 \theta_{13}
      \mbox{sin}^2 \left(\frac{1.27 \Delta m^2 L}{E}\right) \nonumber \\
P(\nu_\mu \rightarrow \nu_\mu) &=& 1 - 4 \cos^2 \theta_{13} \mbox{sin}^2 \theta_{23}  
       ( 1-\cos^2 \theta_{13} \mbox{sin}^2 \theta_{23}) \nonumber \\ 
       && \times \mbox{sin}^2 \left(\frac{1.27 \Delta m^2 L}{E} \right).
\label{eqn:osc-vacuum}
\end{eqnarray}
%
Under this framework the three-neutrino oscillation probability in constant density matter may be written~\cite{Giunti98} as 
%
\begin{equation}
P(\nu_{\mu}  \leftrightarrow \nu_{e} )  =  \mbox{sin}^{2}\theta_{23} \mbox{sin}^{2} 2 \theta_{13}^{M}
                                          \mbox{sin}^{2} \left( \frac{ 1.27 \Delta m^{2}_{M} L   }{ E } \right).
\label{q13oscprob}
\end{equation}

\noindent The matter modified mixing parameters are 

%
\begin{eqnarray}
\Delta m^{2}_{M}                 & = &\Delta m^{2} \sqrt{ \mbox{sin}^{2} 2\theta_{13}
                                     + ( \Gamma - \mbox{cos}\, 2\theta_{13} )^{2} } \\ 
\mbox{sin}^{2} 2 \theta_{13}^{M} & = & \frac{ \mbox{sin}^{2} 2 \theta_{13}  }
                                       { \mbox{sin}^{2}2\theta_{13} + 
                                       ( \Gamma - \mbox{cos}\, 2\theta_{13} )^{2} }, \nonumber
\label{MatterVariables}
\end{eqnarray}
\noindent where $\Gamma = \pm 2 \sqrt{2} G_{f} n_{e} E / \Delta m^{2}$, $G_{f}$ is the Fermi constant, $n_{e}$ is the 
local electron density and the plus (minus) sign specifies neutrinos (anti-neutrinos). 
Resonant enhancement of the oscillation probability occurs when $ |\Gamma| = \mbox{cos}\,2 \theta_{13}$ and holds 
for either neutrinos or anti-neutrinos, depending on the mass hierarchy. Further, when $\theta_{13} = 0$  there 
is no enhancement. 

Oscillation probabilities for neutrinos traversing the Earth appear in the left panel of Fig.~\ref{fig:3fprob}.
For $\Delta m^{2} \sim 2\times 10^{-3}~\mbox{eV}^{2}$ this resonance occurs in the 
2-10 GeV region and its strength increases with \tonethree 
reaching $\sim 40\%$ conversion probability near the Chooz limit. Under these conditions, the 
primary signature in the atmospheric neutrino sample at Super-K is an increased rate 
of high energy upward-going \mbox{$e$-like} events.  The right panel of the figure shows
the $\nu_{e}$ flux ratio at SK oscillated with $\theta_{13}$ at the Chooz limit relative 
to that oscillated at $\theta_{13} = 0.$ Additional $\theta_{13}$-induced effects on muon event rates are expected, 
but are generally much smaller. For large values of \tonethree an expected $\sim 20\%$ increase in the multi-ring 
\mbox{$e$-like} event (see below) rate would be accompanied by a $\sim 5\%$ change in similarly energetic 
muon-like~(\mbox{$\mu$-like})~samples. 

Including solar oscillation terms changes the oscillation probability in the 
resonance region by less than 5\%, supporting our assumption of the ``one mass scale dominant'' framework. 
Their inclusion as an additional scanning parameter also introduces a large computational burden in the $\theta_{13}$ analysis and further motivates a separate $\theta_{23}$ octant analysis.   

  Oscillation probabilities in both analyses are computed using a numerical technique~\cite{Barger80}. Probabilities inside the Earth are computed using 
a piecewise constant radial matter density profile constructed as the median density
in each of the dominant regions of the PREM~\cite{PREM} model: 
inner core $(   0  \leq r < 1220 \mbox{km})$ 13.0 $\mbox{g/cm}^{3}$,
outer core $(1220  \leq r < 3480 \mbox{km})$ 11.3 $\mbox{g/cm}^{3}$,
mantle     $(3480  \leq r < 5701 \mbox{km})$  5.0 $\mbox{g/cm}^{3}$, and
the crust  $(5701  \leq r < 6371 \mbox{km})$  3.3 $\mbox{g/cm}^{3}$.
Transition amplitudes are computed across each layer a neutrino traverses
and the product of these together with the amplitude for crossing the Earth's
atmosphere is used to compute the final oscillation probability. 
The difference in the obtained probabilities using this simplified model compared to 
the more expansive PREM model have a negligible impact on the final analysis results
after incorporating detector resolution effects.

\begin{figure*}[ht]
  \begin{minipage}{6.5in}
    \includegraphics[width=6.0in,type=pdf,ext=.pdf,read=.pdf]{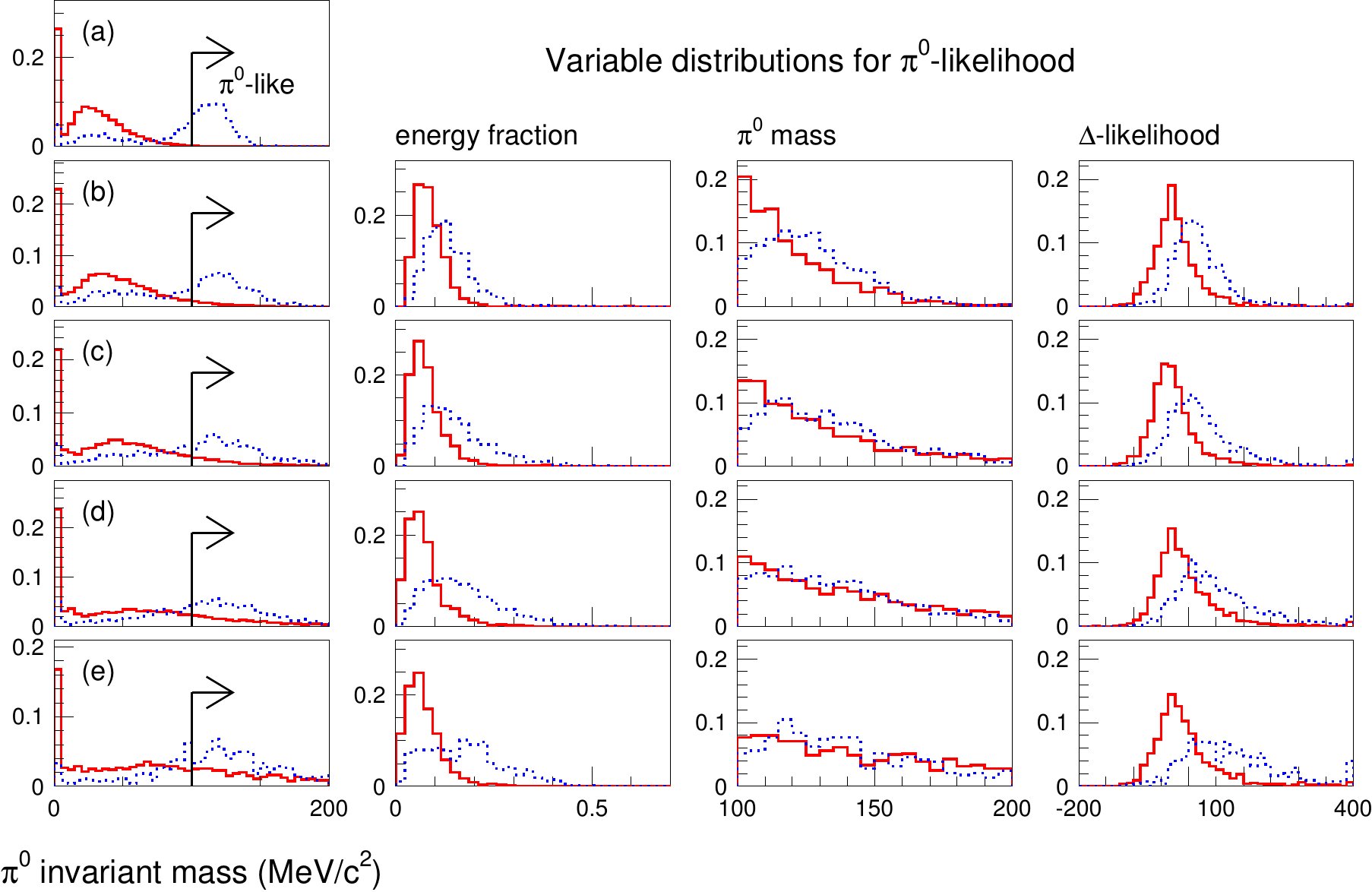}
     \caption{ 
        (color online).
        The distributions used in the $\pi^{0}$ selection for five momentum regions:
	(a)$P_e <$ 250 MeV/c, (b)250 MeV/c $\leq P_e <$ 400 MeV/c, 
	(c)400 MeV/c $\leq P_e <$ 630 MeV/c, (d)630 MeV/c $\leq P_e <$ 1000 MeV/c and (e)1000 MeV/c $\leq P_e$.
	Solid (dashed) lines represent CCQE (NC) events in the FC sub-GeV single-ring \mbox{$e$-like} Monte Carlo.
	Events with an invariant mass above $100~\mbox{MeV/c}^2$ are selected as $\pi^0$-like.
	To separate \mbox{$\pi^0$-like} and electron-like more efficiently, an additional likelihood selection is applied 
	for events with momentum above 250 MeV/c.
	The distributions of the three likelihood variables are shown:
	the fraction of energy carried by the second fitted ring ($E_2/(E_1+E_2)$), 
	the $\pi^0$ mass and $\Delta$-likelihood (described in the text).
	All distributions have been normalized to unit area.
      }
  \label{fig:pi0sel}
  \end{minipage}
\end{figure*}

\section{Data Sample}
\label{sec:datasample}

\superk is a cylindrical 50 kton water Cherenkov detector situated at a depth of 
2700 meters water equivalent. The detector volume is optically separated into 
an inner volume (ID) and an outer veto region (OD). During the 
SK-I (SK-II) periods the ID was instrumented with 11,146 (5,182) inward-facing 20-inch 
photomultiplier tubes (PMTs) and the OD with 1,885 outward-facing 8-inch PMTs.
In SK-III there were 11,129 ID PMTs. 
Since SK-II, the ID PMTs have been encased in fiber-reinforced plastic shells 
with acrylic covers to prevent chain reactions within the detector in the event of a PMT implosion. 
A more detailed description of the detector may be found in~\cite{fukuda:2002uc}.

In this paper, atmospheric neutrino events are organized into three classes:
fully contained (FC), partially contained (PC), and upward-going muons (\upmu).
Events which deposit all of their Cherenkov light in the ID are classified
as FC, while events that originate in the ID but have an exiting particle depositing
energy in the OD are considered PC. Neutrino interactions occurring in the rock beneath 
the detector which produce muons that traverse the detector
(through-going) or stop in 
the detector (stopping) are classified as \upmu events. Data accrued 
in the five years spanning the SK-I run period starting in 1996 correspond 
to 1489 live-days of FC and PC events with 1646 \upmu live-days. SK-II 
data were taken between December 2002 and October 2005 and represent
799 (518) live-days of FC and PC events and 828 live-days of \upmu events.
SK-III data were taken between December 2005 and June 2007 where the FC and PC livetime 
was 518 days and that for \upmu was 635.
The difference of livetimes between FC/PC and \upmu is due to the insensitivity
of the \upmu reduction to noise such as ``flasher'' PMTs. Such noise 
may be misconstrued as real FC/PC events so in those reductions data 
surrounding these events are rejected. 

%
\begin{table*}[htbp]
\begin{center}
\begin{tabular}{ll.....}
\hline
\hline
                 &                 & \multicolumn{3}{c}{FC sub-GeV single-ring $e$-like}   
                                &  & \multicolumn{1}{c}{FC sub-GeV} \\
                 &                 & \multicolumn{1}{c}{0-decay}         & \multicolumn{1}{c}{1-decay} & 
                                     \multicolumn{1}{c}{$\pi^0$-like} &  & \multicolumn{1}{c}{single-ring $e$-like}  \\
\multicolumn{2}{l}{MC events}      & \multicolumn{1}{r}{2663.2}          & \multicolumn{1}{r}{210.9}  
                                   & \multicolumn{1}{c}{191.8}        &  & \multicolumn{1}{c}{2996.4} \\ 
                 & Q.E.            & 77.7\,\%            & 3.8\,\%       & 10.6\,\%         &  & 70.6\,\%  \\ 
CC               & single meson    & 12.4\,\%            & 50.3\,\%      & 7.0\,\%          &  & 15.2\,\%  \\
$\nu_{e}+\bar\nu_{e}$ & multi $\pi$& 1.0\,\%             & 9.7\,\%       & 1.8\,\%          &  & 1.7\,\%   \\
                 & coherent $\pi$  & 1.3\,\%             & 8.5\,\%       & 0.5\,\%          &  & 1.7\,\%   \\
\multicolumn{2}{l}{CC $\nu_{\mu}+\bar\nu_{\mu}$}& 0.6\,\%  & 15.2\,\%  & 7.0\,\%          &  & 2.0\,\%  \\ 
\multicolumn{2}{l}{NC}           & 6.8\,\%             & 11.2\,\%      & 72.0\,\%         &  & 8.7\,\% \\
\\
                 &                 & \multicolumn{3}{c}{FC sub-GeV single-ring $\mu$-like}     
                                &  & \multicolumn{1}{c}{FC sub-GeV} \\
                 &                 & \multicolumn{1}{c}{0-decay}    & \multicolumn{1}{c}{1-decay} & 
                                     \multicolumn{1}{c}{2-decay} &  & \multicolumn{1}{c}{single-ring $\mu$-like}  \\
\multicolumn{2}{l}{MC events} 
                                   & \multicolumn{1}{r}{1412.4}     & \multicolumn{1}{r}{2745.4}  
                                   & \multicolumn{1}{c}{164.3}   &  & \multicolumn{1}{c}{4297.8} \\ 
                 & Q.E.            & 71.3\,\%            & 78.5\,\%       & 5.8\,\%          &  & 74.7\,\%  \\ 
CC               & single meson    & 12.9\,\%            & 15.5\,\%       & 65.7\,\%         &  & 16.7\,\%  \\
$\nu_{\mu}+\bar\nu_{\mu}$ & multi $\pi$& 1.1\,\%         & 1.5\,\%        & 14.9\,\%         &  & 1.9\,\%   \\
                 & coherent $\pi$  & 0.8\,\%             & 1.5\,\%        & 8.6\,\%          &  & 1.6\,\%   \\
\multicolumn{2}{l}{CC $\nu_{e}+\bar\nu_{e}$}& 1.8\,\%    & <0.1\,\%       & <0.1\,\%         &  & 0.7\,\%  \\ 
\multicolumn{2}{l}{NC}           & 11.8\,\%              & 2.6\,\%        & 3.3\,\%          &  & 4.3\,\% \\
\hline
\hline
\end{tabular}

\caption{ The number of FC sub-GeV MC events and their fractional composition
          by neutrino interaction mode in SK-I. The upper(lower) table shows the \mbox{$e$-like} (\mbox{$\mu$-like}) sample.
          The left (right) side of the table shows the result after (before) separation into sub-samples. 
          After separation, the CCQE purity is increased and the NC backgrounds are reduced 
          in the 0-decay \mbox{$e$-like} and 1-decay \mbox{$\mu$-like} sub-samples.
         }
\label{table:fcfrac}
\end{center}
\end{table*}
 
Fully contained events are further divided into sub-GeV and multi-GeV sub-samples 
based on visible energy, \evis. Events with \evis $ < 1.33 $ GeV are considered sub-GeV.
The number of reconstructed Cherenkov rings in an event is also used to separate 
these samples into single- and multi-ring sub-samples. 
Single-ring events are classified into $\mu$-like and $e$-like samples 
by the ring pattern.
For multi-ring samples, the most energetic ring is used to classify the event type.
Partially contained events are classified as ``OD stopping'' or ``OD through-going'' based on their energy deposition 
in the OD~\cite{ashie:2004mr}. Similarly, \upmu events that traverse the detector are 
separated into ``showering'' and ``non-showering'' based on the method described in~\cite{Desai08}
while those that enter and stop within the detector are classified as ``stopping.''
These samples are defined for all of the SK run periods. 
To enhance each analysis' sensitivity to the desired oscillation effect, the FC samples have
been further divided as outlined below. However, all of the data samples are used in both analyses.

  Several improvements to the reconstruction and Monte Carlo since earlier publications~\cite{ashie:2005ik,Hosaka:2006zd} are incorporated in this paper. The ring counting likelihood has been updated to improve separation between the single-ring and multi-ring samples. 
Additionally, the neutrino interaction generator has been updated to include lepton mass effects in charged current (CC) interactions~\cite{Berger2007,Kuzmin2003}. An axial vector mass of 1.2 GeV has been used for quasi-elastic and single meson production processes and cross sections for deep inelastic scattering are computed based on the GRV98 parton distribution functions~\cite{GRV98}. 
The atmospheric neutrino flux is taken from~\cite{honda}. More detailed information on the MC simulation, event generator, and event reconstruction is presented in~\cite{ashie:2005ik}.

\subsection{Additional sample selection for the $\theta_{23}$ octant analysis}

  To increase the purity of the interaction mode,  FC sub-GeV single-ring
events are separated into sub-samples based on their number of decay electrons 
and how \mbox{$\pi^{0}$-like} they are.

The FC sub-GeV single-ring \mbox{$e$-like} sample contains background events which are 
mainly neutral-current (NC) $\pi^{0}$ events where one of the two $\gamma$ rays
from the $\pi^{0}$ decay has been missed by the event reconstruction. The 
electromagnetic shower from the $\gamma$ gives a light pattern similar to that 
of an electron and results in an electron-like classification. To reduce this 
type of background, a specialized $\pi^{0}$ fitter is used~\cite{thesis:tomasz}. This fitter 
enforces a second ring on the data and then predicts a light pattern that would result 
from $\gamma$ rays propagating through the tank with the direction and vertex of 
the fitted rings. The intensity of each fitted ring as well as its direction are varied 
until the predicted light pattern best agrees with the observed one. 
Since the interaction mode of interest, charged-current quasi-elastic (CCQE), creates 
only one light-emitting particle, constructing the invariant mass for the two fitted
rings provides some separation between CCQE and NC events.
The left five panels in Fig.~\ref{fig:pi0sel} show the invariant mass distributions from this $\pi^0$ fitter
for CCQE and NC events in the FC sub-GeV single-ring \mbox{$e$-like} Monte Carlo in five energy regions.
Neutral current events tend to form a peak close to the $\pi^{0}$ mass whereas CCQE do not.
For events with electron momentum below 250 MeV/c, a cut at~$100~\mbox{MeV/c}^2$ is used to create 
a \mbox{$\pi^{0}$-like} sample. This cut, however, is not sufficient for higher electron momenta so 
an additional likelihood selection is used, incorporating three variables: 
the $\pi^{0}$ invariant mass distribution, the fraction of the event's reconstructed momentum 
carried by the second ring, and
the difference of two likelihood variables which result from a $\pi^0$-fit and electron-fit.
The distribution of these variables is shown in Fig.~\ref{fig:pi0sel}.
The \mbox{$\pi^0$-like} selection likelihood functions are defined as, 
%
\begin{equation}
\mathcal{L} = \displaystyle\sum_{i=1}^{3} \mbox{log}(\Gamma^{S}_{i}( x_{i} )) -\mbox{log}(\Gamma^{B}_{i}( x_{i} )),
\label{pi0likelihood}
\end{equation}
where $\Gamma^{S}_{i}( x_{i} )$($\Gamma^{B}_{i}( x_{i} )$) represents the CCQE(NC) events' probability distribution function (PDF) 
for the $i^{th}$ variable with observable $x_{i}$. 

  After separating the \mbox{$\pi^0$-like} sample, the remaining \mbox{$e$-like} events are divided into two
categories, 0-decay which has no decay electrons and 1-decay which has one or more decay electrons.
Since $\nu_e$ CCQE events are not expected to produce decay electrons, there is a large fraction 
of CCQE interactions in the 0-decay sample.
For the FC sub-GeV single-ring \mbox{$\mu$-like} sample, there are three categories using the number
of decay electrons: 0-decay, 1-decay, and 2-decay, corresponding to the number of decay electrons reconstructed in the event.
Since these CCQE events produce a muon they are expected to have at least one decay electron.
Details of the event composition by interaction mode after these event selections are shown in 
Tbl.~\ref{table:fcfrac}. The fraction of CCQE events is increased in the 0-decay \mbox{$e$-like} and 
1-decay \mbox{$\mu$-like} samples so they should improve
sensitivity to changes in the sub-GeV flux induced by solar oscillations.

\subsection{Additional sample selection for the $\theta_{13}$ analysis }

To improve sensitivity to \nue appearance induced by non-zero \tonethree,
an enhanced FC multi-GeV multi-ring electron-like sample is created. 
The selection is based on a likelihood method~\cite{Hosaka:2006zd}
for SK-I and is extended in this analysis to SK-II and SK-III. 
The likelihood functions have been rebuilt using 100 years of MC incorporating 
recent improvements to the SK event reconstruction.  Accordingly, the event populations of 
the SK-I sample here differ from those in the reference.

\begin{figure*}[htbp]
  \begin{minipage}{6.5in}
    \includegraphics[height=4.0in,width=6.0in,type=pdf,ext=.pdf,read=.pdf]{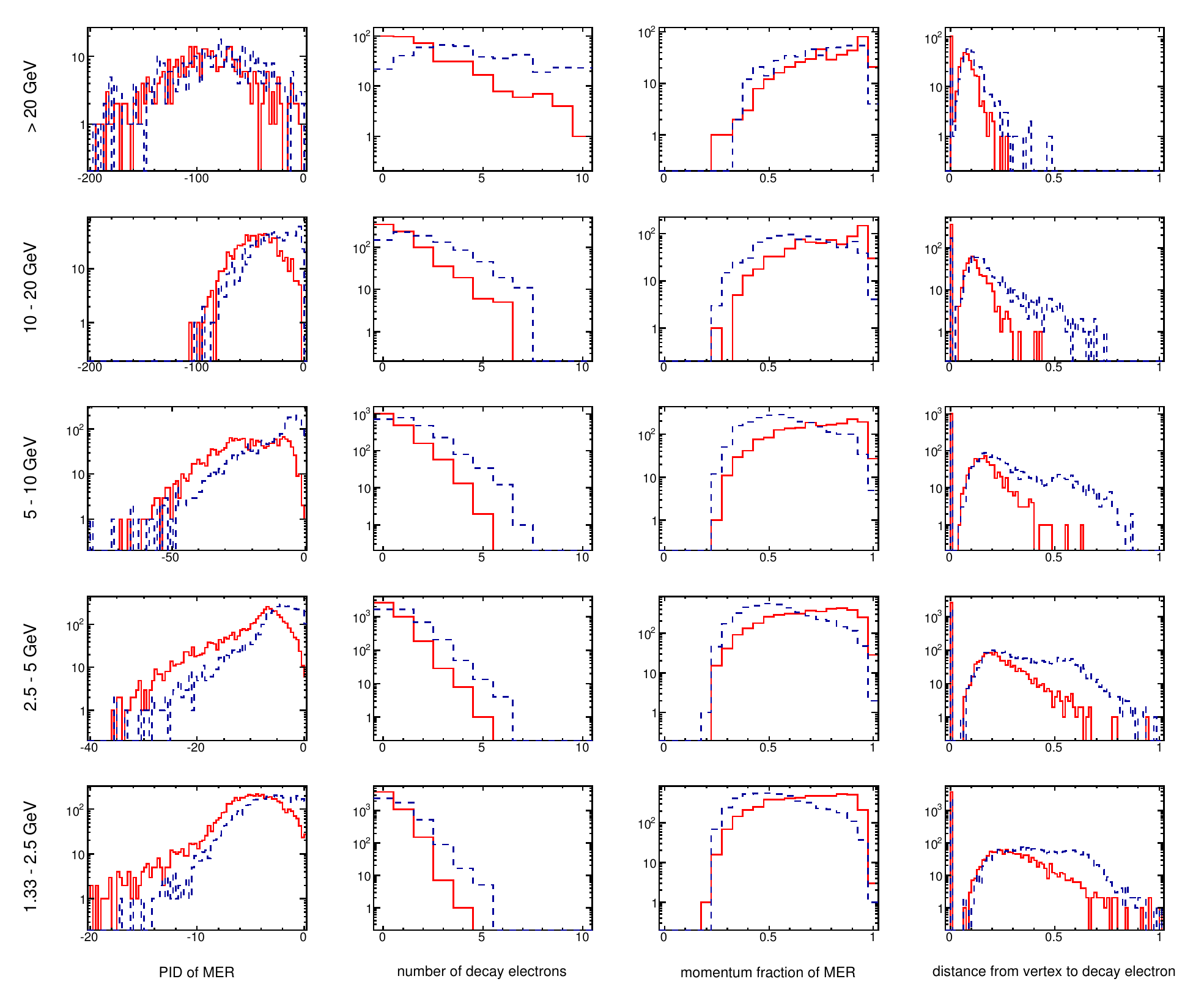}
     \caption{ 
        (color online).
                Variables used in the likelihood definition to create the SK-I
                multi-GeV multi-ring \mbox{$e$-like} sample. The energy bins correspond to the 
                the most energetic ring (MER) and the distributions have been scaled to the SK-I livetime.
                 Signal events (CC $\nu_{e} + \bar \nu_{e}$)
                are shown as the solid line and background events (CC $\nu_{\mu} + \bar \nu_{\mu}$ 
		and NC) are shown as the dashed line. 
                The shapes of these distributions do not differ appreciably among SK-I, SK-II, and SK-III. 
             }
  \label{fig:mgmrelikelihood}
  \end{minipage}
\end{figure*}

The MC is divided into five energy bins, 1.33-2.5 GeV, 2.5-5 GeV, 5-10 GeV, 10-20 GeV, and $> 20$ GeV
and PDFs for each bin are constructed using events whose most energetic ring has been 
reconstructed as electron-like. Four observables are used in the event selection:
the number of decay electrons in the event, 
the maximum distance between the neutrino vertex and any
muon decay electrons, the fraction of momentum carried by the event's most energetic ring,
and the PID likelihood value of that ring. The final likelihood functions are defined as, 

%
\begin{equation}
\mathcal{L}_{j} = \displaystyle\sum_{i=1}^{4} \mbox{log}(\Gamma^{S}_{i}( x_{i} )) -\mbox{log}(\Gamma^{B}_{i}(x_{i})),
\label{mrelikelihood}
\end{equation}

\noindent where $\Gamma_{i}$ represents the PDF for the $i^{th}$ observable and $x_{i}$ is the observable's
measured value. The superscripts $S$ and $B$ label the signal and background PDFs, respectively. The 
index $j$ specifies the likelihood corresponding to one of the five energy bins considered. 
In selecting electron neutrino events, the signal is taken to be CC $\nu_{e} + \bar \nu_{e}$,
while the background is composed of both CC $\nu_{\mu} + \bar \nu_{\mu}$ and NC events. 
An event makes it into the final multi-GeV multi-ring sample if it passes all cuts in the FC reduction,
if the event's most energetic ring is electron-like, and if $\mathcal{L}_{j} > 0$.
Distributions of the likelihood variables for signal and background events appear in Fig.~\ref{fig:mgmrelikelihood}. 
Decay electrons are produced in the signal sample through the decay chain of pions produced in these events.
However, lacking an exiting muon at the interaction vertex, fewer decay electrons are expected in the signal sample. 
Similarly, the maximum distance to a decay electron in the CC component of the background is expected to be larger 
due to the presence of energetic muons. 
The distribution of the momentum fraction carried by the most energetic ring tends to peak towards higher values 
for signal events where the outgoing electron has been correctly identified as electron-like.
Background events, on the other hand, tend to peak at lower momentum fractions where the most energetic ring comes from 
a meson or muon that has been misidentified as electron-like. 
Applying the likelihood improves the signal purity from 53\% to 74\% in SK-I with 16\% of the sample coming 
from NC events. 
Table~\ref{tbl:MREpurity} shows the event compositions of the multi-GeV \mbox{$e$-like} sample after this selection for SK-I, SK-II, and SK-III.

%
\begin{table*}[htbp]
   \begin{center}
      \begin{tabular}{l........}
      \hline
      \hline
             &\multicolumn{2}{c}{CC $\nu_{e} + \bar \nu_{e}$} 
             &\multicolumn{2}{c}{CC $\nu_{\mu} + \bar \nu_{\mu}$} 
             &\multicolumn{2}{c}{NC} 
             &\multicolumn{2}{c}{ Total } \\

             &\multicolumn{1}{c}{ No $\mathcal{L}$ }  
             &\multicolumn{1}{c}{ $\mathcal{L}$ }     
             &\multicolumn{1}{c}{ No $\mathcal{L}$ }  
             &\multicolumn{1}{c}{ $\mathcal{L}$ }     
             &\multicolumn{1}{c}{ No $\mathcal{L}$ }  
             &\multicolumn{1}{c}{ $\mathcal{L}$ }     
             &\multicolumn{1}{c}{ No $\mathcal{L}$ }  
             &\multicolumn{1}{c}{ $\mathcal{L}$ }    \\
      \hline

       SK-I            & 472.1 &331.0 & 201.7 &39.2 & 218.1 & 74.4 & 891.9 & 444.6 \\
       Percentage (\%) &  53.0 & 74.5 &  22.6 & 8.8 &  24.5 & 16.7 & 100.0 & 100.0 \\
       \\
       SK-II           & 253.4 &178.0 & 110.7 &23.4 & 119.1 & 42.8 & 483.2 & 244.2 \\
       Percentage (\%) &  52.4 & 72.9 &  23.0 & 9.6 &  24.6 & 17.5 & 100.0 & 100.0 \\
       \\
       SK-III          & 157.6 &112.8 &  72.7 &15.1 &  74.0 & 26.1 & 304.3 & 154.0 \\
       Percentage (\%) &  51.8 & 73.2 &  23.9 & 9.8 &  24.3 & 17.0 & 100.0 & 100.0 \\
      \hline
      \hline
      \end{tabular}
      \caption{ 
         The expected number of events for each interaction component
         of the multi-ring multi-GeV \mbox{$e$-like} sample 
         before (No $\mathcal{L}$) and after ($\mathcal{L}$)
         likelihood selection for the SK-I, SK-II, and SK-III MC scaled to 1489,
         798, and 518 livetime days, respectively. 
         Two flavor neutrino oscillations
         $\nu_{\mu} \leftrightarrow \nu_{\tau}$ have been assumed with
         $\Delta m^{2} = 2.1 \times 10^{-3}~\mbox{eV}^{2}$ and
         $\mbox{sin}^{2}2\theta = 1.0$.
      \label{tbl:MREpurity}
      }
   \end{center}
\end{table*}

  A summary of all atmospheric neutrino event samples used in this paper is 
shown in Tbl.~\ref{table:evsummary}. 

\begin{table*}[htbp]
\begin{center}
\begin{tabular}{l.........}
\hline \hline
        & \multicolumn{3}{c}{SK-I}                
        & \multicolumn{3}{c}{SK-II}               
        & \multicolumn{3}{c}{SK-III}  \\           

        & \multicolumn{1}{c}{Data}                  
        & \multicolumn{1}{c}{MC}
        & \multicolumn{1}{c}{(osc.)}
        & \multicolumn{1}{c}{Data}                  
        & \multicolumn{1}{c}{MC}
        & \multicolumn{1}{c}{(osc.)}
        & \multicolumn{1}{c}{Data}                  
        & \multicolumn{1}{c}{MC}
        & \multicolumn{1}{c}{(osc.)} \\
\hline
\multicolumn{10}{l}{FC Sub-GeV} \\
\multicolumn{10}{l}{single-ring} \\
\multicolumn{10}{l}{~~~$e$-like} \\
~~~~~~~~0-decay           &  2984 & 2655.9  &(2652.4) & 1605 & 1405.8 &(1403.4) & 1098 & 935.7 &(934.7)\\ 
~~~~~~~~1-decay           &   275 & 204.4  &(194.3)   & 155  & 113.5 &(107.1)   & 106 & 69.6 &(66.7)\\
~~~~~~~~$\pi^0$-like      &   167 & 159.1  &(155.2)   & 81   & 81.3 &(79.1)     & 46 & 45.6 &(44.4)\\
\multicolumn{10}{l}{~~~$\mu$-like} \\
~~~~~~~~0-decay           &  1036 & 1385.6  &(973.0)  & 563  & 765.9 &(537.2)  & 346 & 497.5 &(350.4)\\ 
~~~~~~~~1-decay           &  2035 & 2760.6  &(1846.8) & 1043 & 1429.4 &(957.3) & 759 & 999.8 &(668.3)\\
~~~~~~~~2-decay           &   150 & 163.7  &(114.6)   & 80   & 82.8 &(57.7)    & 61 & 58.5 &(41.0)\\
~~~2-ring $\pi^0$-like    &   497 & 460.0  &(456.1)  & 267  & 237.3 &(235.1) & 178 & 157.8 &(156.5)\\
\\
\multicolumn{10}{l}{FC Multi-GeV} \\
\multicolumn{10}{l}{single-ring} \\
~~~$e$-like             & 829 & 777.8  &(777.7)   & 392 & 409.9 &(411.1) & 282 & 279.3 &(278.4)\\
~~~$\mu$-like           & 694 & 1027.4  &(744.4)  & 394 & 550.2 &(399.0) & 231 & 352.8 &(255.7)\\
\multicolumn{10}{l}{multi-ring} \\
~~~$e$-like             & 433 & 457.9  &(458.9)   & 260 & 252.3 &(251.9) & 149 & 159.2 &(159.1)\\
~~~$\mu$-like           & 617 & 882.4  &(660.9)   & 361 & 459.6 &(344.1) & 226 & 313.8 &(234.4)\\
\\
\multicolumn{10}{l}{PC} \\
~~~OD stopping          & 163 & 222.7  &(167.3)   & 116 & 105.8 &(80.9)  & 63 & 75.1 &(55.7)\\
~~~OD through-going     & 735 & 965.4  &(755.0)   & 314 & 482.5 &(374.7) & 280 & 334.9 &(262.8)\\
\\
\multicolumn{10}{l}{Upward-going muon} \\
~~~stopping            & 435.9  & 701.7  &(419.4) & 207.6  & 355.2 &(212.5) & 193.7 & 273.8 &(163.5)\\
~~~non-showering       & 1577.4 & 1548.0 &(1343.9) & 725.3  & 767.6 &(668.7) & 612.9 & 599.4& (520.0)\\
~~~showering           & 271.6  & 302.7  &(292.2)   & 108.1  & 147.8 &(143.6) & 110 & 116.2 &(112.3)\\
\\
\multicolumn{10}{l}{Reduction Efficiency} \\
~~~FC                  &  & & 97.6\,\%
                    &  & & 99.2\,\% 
                    &  & & 98.5\,\% \\
~~~PC               &  & & 81.0\,\% 
                    &  & & 74.8\,\% 
                    &  & & 88.6\,\% \\
~~~Upward stopping $\mu$  & & & 98.0\,\% 
                          & & & 97.0\,\% 
                          & & & 98.2\,\% \\
~~~Upward through-going $\mu$ & & & 99.4\,\% 
                           & & & 98.1\,\%
                           & & & 99.4\,\% \\
\hline
\hline
\end{tabular}
\caption{Summary of atmospheric neutrino data and MC event samples for FC, PC, and \upmu
in SK-I, SK-II, and SK-III. The FC and PC livetime is 1489\,days in SK-I, 
798\,days in SK-II, and 518\,days in SK-III. The livetime of the \upmu samples is 1645\,days in SK-I, 
827\,days in SK-II, and 635\,days in SK-III. The number of MC events has been normalized 
by the livetime of the data. The oscillated MC has been calculated using two flavor mixing at
$\Delta m^2 =2.1 \times 10^{-3}$ eV$^2$ and sin$^2 2\theta =1.0$.
}
\label{table:evsummary}
\end{center}
\end{table*}

\section{Oscillation Analysis }
\label{sec:analysis}

  The oscillation analyses have been performed using the above data samples.
 Since the physical detector configuration differs between SK-I, SK-II, and SK-III, 
separate 500 years-equivalent MC data sets for each run 
period are used. The data are compared against the MC expectation using a 
``pulled'' $\chi^{2}$~\cite{Lisi02} method based a Poisson probability distribution:

%
\begin{widetext}
\begin{eqnarray}
\chi^{2}  = 2 \displaystyle \sum_{n} \left(  E_{n}( 1 + \displaystyle \sum_{i} f^{i}_{n} \epsilon_{i} )
               -\mathcal{O}_{n}  + \mathcal{O}_{n} \ln \frac{ \mathcal{O}_{n} }{ E_{n}( 1 + \displaystyle \sum_{i} f^{i}_{n} \epsilon_{i} ) } \right)
             + \displaystyle \sum_{i} \left( \frac{ \epsilon_{i} }{ \sigma_{i} } \right)^{2}, 
\label{eq:fullchi}
\end{eqnarray}
\end{widetext}

\noindent 
where $n$ indexes the data bins, $E_{n}$ is the MC expectation, and $\mathcal{O}_{n}$ is the 
number of observed events in the $n^{th}$ bin. Systematic errors are incorporated into the fit 
via the systematic error parameter $\epsilon_{i}$, where $i$ is the systematic error index and 
$f^{i}_{n}$ is the fractional change in the MC expectation in bin $n$ for a 1-sigma change in 
the $i^{th}$ systematic error. The 1-sigma value of a systematic error is labeled as 
$\sigma_{i}$.  Equation~(\ref{eq:fullchi}) is minimized with respect to the $\epsilon_{i}$ at each point in 
 a fit's oscillation parameter space according to $\frac{ \partial \chi^{2}}{ \partial \epsilon_{i}} = 0.$
This derivative yields a set of linear equations in $\epsilon$ that can be solved iteratively~\cite{Lisi02}. 
The best fit point is defined as the global minimum $\chi^{2}$ on the grid of oscillation points.

 To ensure stability of the function in Eq.~(\ref{eq:fullchi}) the binning has been chosen
so that there are at least 6 expected MC events in each bin after scaling to the SK-I livetime.
Data are binned separately for SK-I, SK-II, and SK-III, each with a total of 420 bins.
Both analyses simultaneously fit 16 event samples, including both e-like and $\mu$-like event categories, 
shown in Tbl. \ref{table:evsummary}. 
The samples separated by number of decay electrons are divided into 5 momentum and 10 zenith angle bins for the 
0-decay \mbox{$e$-like}, 0-decay and 1-decay \mbox{$\mu$-like} samples, and 1 zenith bin otherwise. The remaining FC events are divided among 14
momentum bins, PC events into a total of 6 bins, and all upward through-going muon 
samples have one momentum bin each.
The upward-stopping muon samples have been divided into three momentum bins.
All of these samples are further divided into 10 evenly-spaced zenith angle bins.
FC and PC events range from $ -1 \le \mbox{cos} \Theta \le 1$ and \upmu events are 
binned from  $ -1 \le \mbox{cos} \Theta \le 0$.

Both analyses consider 120 sources of systematic uncertainty. These systematic errors 
are separated into two categories, those that are common to all of the SK run periods 
and those that differ. Errors that are classified as common are related to uncertainties in 
the atmospheric neutrino flux, neutrino interaction cross sections, and particle production
within nuclei. Systematic errors that are independent for SK-I, SK-II, and SK-III represent 
uncertainties related to the detector performance in each era. Particle reconstruction
and identification uncertainties, as well as energy scale and fiducial volume uncertainties,
differ for SK-I, SK-II, and SK-III because of their different geometries. These systematics are 
therefore considered as separate sources of uncertainty. The effect of the systematic
uncertainties are introduced by the coefficients $f^{i}_{n}$ which are computed for every bin
and error in the analysis. For common systematic uncertainties there is a coefficient
for every bin in the analysis. On the other hand, independent systematic errors 
specific to SK-I (II, III) have non-zero coefficients for the SK-I (II, III) analysis bins
and are zero otherwise. 
Tables~\ref{tbl:systcommonflux} and  ~\ref{tbl:systcommonint} list the 33 common errors separated 
into neutrino flux and interaction-related systematics, respectively. 
Table  ~\ref{tbl:systseparate} lists the $29 \times 3$  independent systematic errors and all three tables 
show errors with their fitted value, $\epsilon_{i}$, from the $\theta_{13}$ search, together with their uncertainty. 
More information about these systematic errors is presented in a previous analysis~\cite{ashie:2005ik}. 

To prevent instabilities in the $\chi^{2}$ value resulting from the low statistics data in later SK run periods, 
the SK-II and SK-III bins are merged with those of SK-I. 
In the minimization of the function in Eq.~(\ref{eq:fullchi}) the following changes are made: 
\begin{eqnarray}
  \mathcal{O}_{n} &\rightarrow & \displaystyle \sum_{i} \mathcal{O}_{n}^{SKi} \nonumber \\ 
  E_{n}( 1 + \displaystyle \sum_{j} f^{j}_{n} \epsilon_{j} ) 
                  &\rightarrow & \displaystyle \sum_{i} E_{n}^{SKi} 
                   ( 1 + \displaystyle \sum_{j} f^{j}_{n} \epsilon_{j} ).  \nonumber   
\label{eqn:mergedchi2}
\end{eqnarray}
Since the systematic error coefficients are computed for separate SK-I, SK-II, and SK-III bins as discussed 
above, merging in this way preserves the effect of the systematic errors specific to each detector geometry.
Using this method, the final $\chi^{2}$ is taken over 420 merged bins.
\begin{table*}[htbp]
   \begin{center}
      \begin{tabular}{lccccc}
      \hline
      \hline
      \\[0.1mm]
      SK-I+II+III  & $\mbox{sin}^{2} \theta_{13}$ 90\% C.L. & $\mbox{sin}^{2} \theta_{13}$ & 
                   $\Delta m^{2} [\mbox{eV}^2]$ & $\mbox{sin}^{2} \theta_{23}$ & $\chi^{2}$/D.O.F \\
      \hline
      \\[0.1mm]
       Normal Hierarchy&   $< 0.04$  &   0.00    &   $2.1\times 10^{-3}$  &  0.50  & 468.7 / 417  \\
       Inverted Hierarchy& $< 0.09$  &   0.006   &   $2.1\times 10^{-3}$  &  0.53  & 468.4 / 417  \\
      \hline
      \hline
      \end{tabular}
      \caption{
          Best fit information for fits to the SK-I+II+III data for both hierarchies in the \tonethree analysis. 
          The limit on $\mbox{sin}^{2} \theta_{13}$ is the C.L. in one-dimension at 90\% and the 
          corresponding bounds on $\Delta m^{2}$ are
          $1.9 (1.7) \times 10^{-3} < \Delta m^{2} < 2.6 (2.7) \times 10^{-3} \mbox{eV}^{2}$ in the
          normal (inverted) hierarchy. 
         \label{tbl:fits}
         }
   \end{center}
\end{table*}
%
%
%
\subsection{\ttwothree octant analysis}
    
   In the search for $\theta_{23} \ne \pi /4$, two fits are performed to the data to extract a 
constraint on $\mbox{sin}^{2}\theta_{23}$ assuming $\theta_{13} = 0$. 
The first (solar-off) is done over the two-dimensional space of $\Delta m^{2}_{23}$ and $\mbox{sin}^{2}\theta_{23}$ 
(41 points each of 
$\Delta m^{2}_{23}$ in $[1.0, 6.3]\times 10^{-3} \mbox{eV}^{2}$ and 
$\mbox{sin}^2 \theta_{23}$ in $[ 0.3, 0.7 ]$) 
and is compared to a second (solar-on) fit, expanded to four dimensions including the solar 
oscillation parameters $\Delta m^{2}_{12}$ and $\mbox{sin}^{2}\theta_{12}$ 
(fit over
4 points of $\Delta m^{2}_{12}$ in $ [7.41, 7.94]\times 10^{-5} \mbox{eV}^{2}$ and 
5 points of $\mbox{sin}^2 \theta_{12}$ in $[0.28, 0.36]$). 
This grid of points has been chosen based on a combined fit of the solar neutrino experiment 
and KamLAND data~\cite{Fogli08,Schwetz08}.  
To constrain the fit over the solar parameters, the $\Delta\chi^2_{solar}$ value from this combined analysis 
is then added to that of the fit at each of these grid points.

\begin{figure}[htb]
  \begin{minipage}{3.3in}
    \includegraphics[width=2.8in,type=pdf,ext=.pdf,read=.pdf]{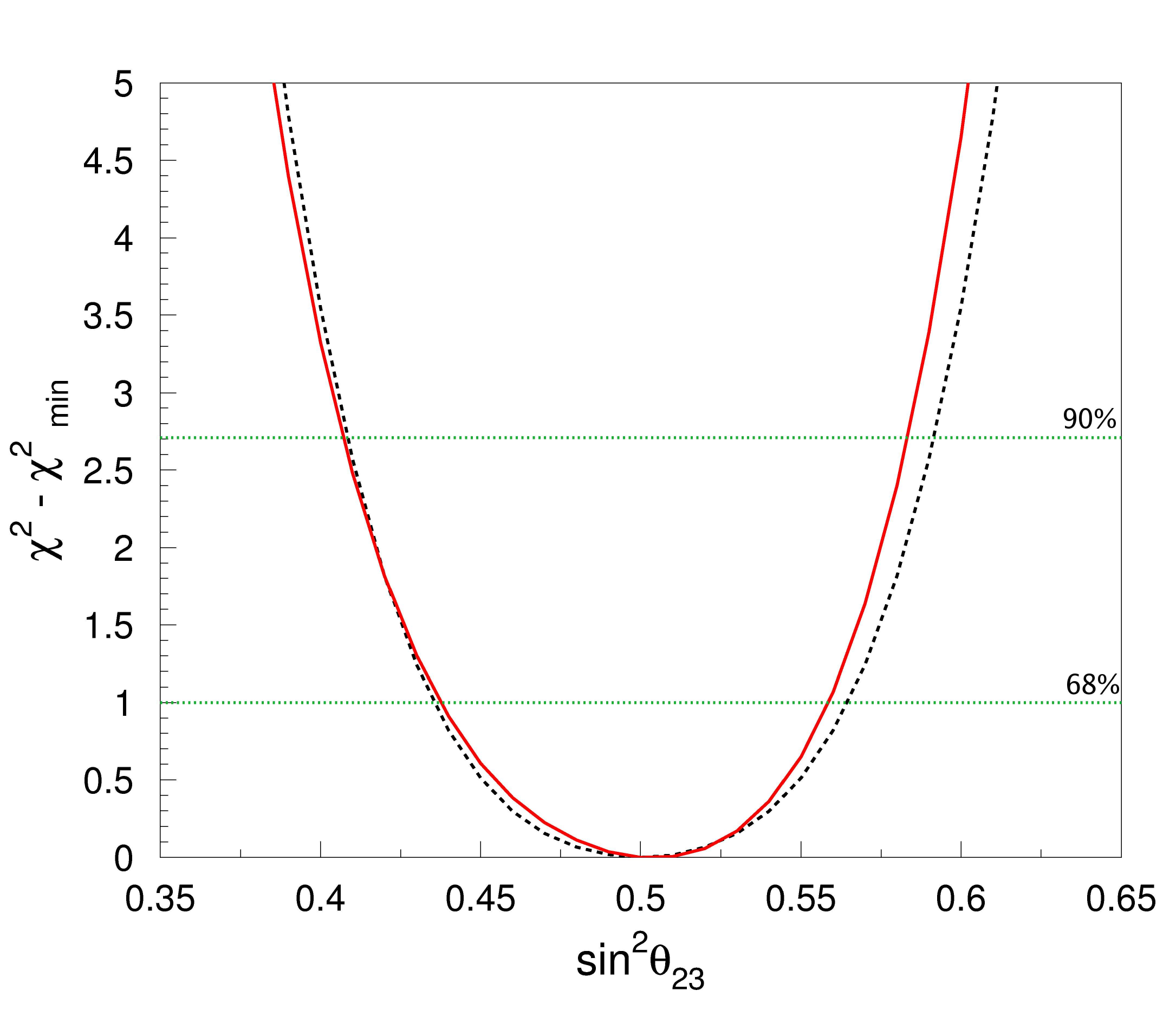}
  \end{minipage}
  \caption{
        (color online).
  $\chi^2$-$\chi^2_{min}$ distribution as a function of sin$^2\theta_{23}$ for oscillations 
  without the 1-2 parameters (dotted line) and with the 1-2 parameters (solid line).
  For each sin$^2\theta_{23}$ point, $\Delta m^2_{23}$ is chosen so that $\chi^2$ is minimized.
  The horizontal line corresponds to the 68\,\%(90\,\%) confidence level which is located at $\chi^2_{min} + 1.0(2.7)$.
     }
\label{fig:sn23_sol}
\end{figure}

\begin{figure}[htbp]
  \begin{minipage}{2.5in}
    \includegraphics[width=2.4in,type=pdf,ext=.pdf,read=.pdf]{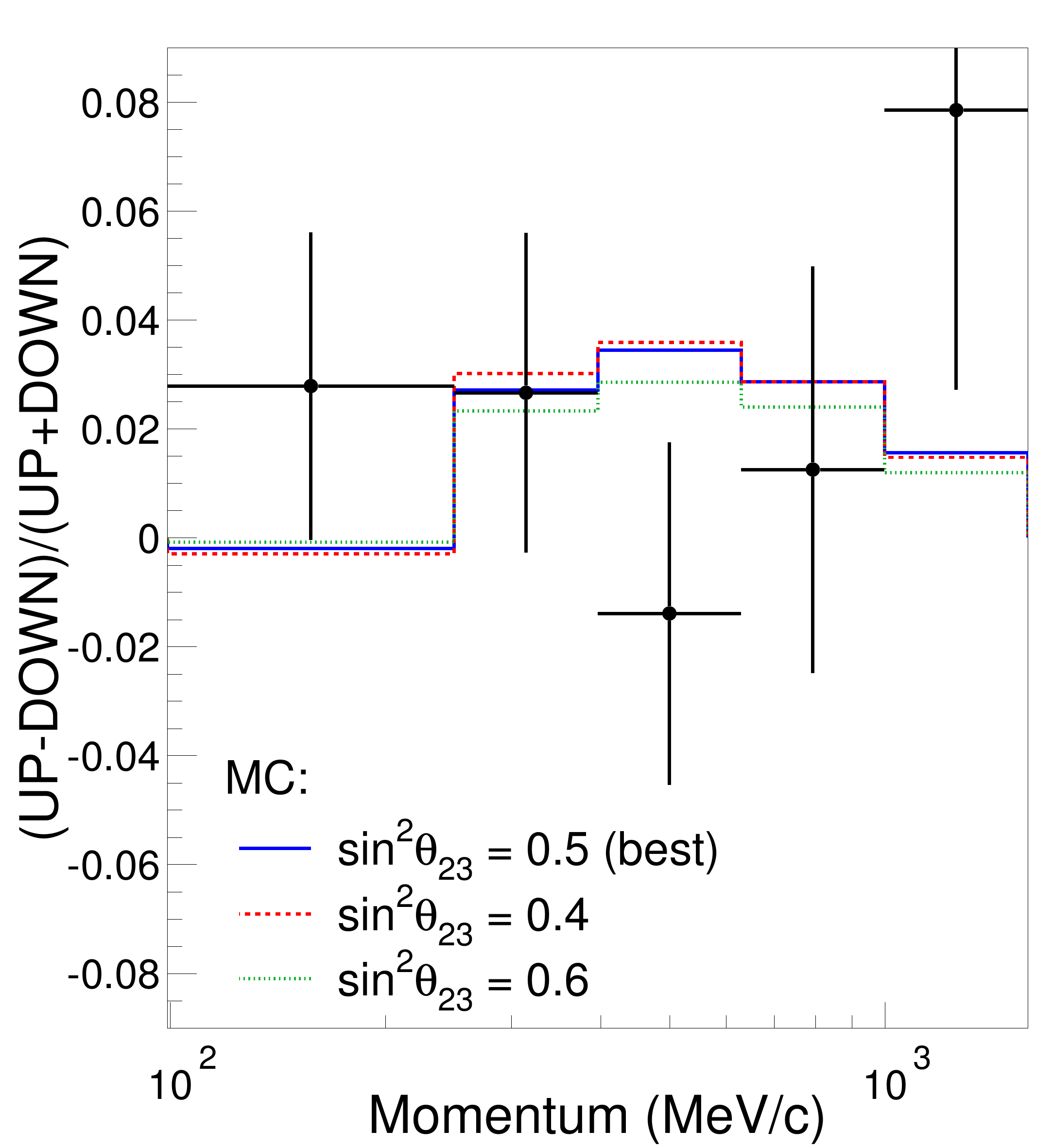}
  \end{minipage}
    \caption{ 
        (color online).
              Asymmetry (Up - Down)/(Up + Down) for the SK-I+II+III single-ring $e$-like 0-decay data set in the octant analysis.
              Up is defined as events with $\mbox{cos}\Theta < - 0.2$ and down as $\mbox{cos}\Theta > 0.2.$
              The solid line represents the MC expectation at the best fit point and the dashed (dotted) line 
              shows the expected asymmetry for $\mbox{sin}^2 \theta_{23} = 0.4 (0.6).$ The error bars are statistical. 
            }
  \label{fig:sk1o2o3q23asymmetry}
\end{figure}

  Figure~\ref{fig:sn23_sol} shows the $\Delta\chi^2$ distributions with and without the solar parameters
as a function of $\mbox{sin}^{2}\theta_{23}$, where $\Delta m^{2}_{12}, \Delta m^{2}_{23}$, 
and $\mbox{sin}^{2}\theta_{12}$ are chosen so that $\chi^2$ is minimized.
The best-fit point with the solar parameters is located at
$\mbox{sin}^{2}\theta_{23}=0.50$, 
$\Delta m^{2}_{23} = 2.1 \times 10^{-3}~\mbox{eV}^{2},$ 
$\mbox{sin}^{2}\theta_{12}=0.30$, and 
$\Delta m^{2}_{12} = 7.59 \times 10^{-5}~\mbox{eV}^{2},$ 
with a minimum $\chi^{2}$ of 470.6/416 d.o.f., 
while that without the solar parameters is 
$\mbox{sin}^{2}\theta_{23}=0.50$, 
$\Delta m^{2}_{23} = 2.1 \times 10^{-3}~\mbox{eV}^{2},$ 
with a minimum $\chi^{2}$ of 469.6/418 d.o.f. 
No significant deviation of $\mbox{sin}^{2}\theta_{23}$ from $\pi/4$ is seen with the addition of 
solar terms to the analysis but they do give rise to the asymmetric shape seen in the $\chi^2$ distribution.
Including the solar terms constrains the measurement of $\mbox{sin}^{2}\theta_{23}$ at the 68 (90)\% C.L. to 
 $0.438 (0.407) < \mbox{sin}^{2}\theta_{23} < 0.558 (0.583).$ 
The up-down asymmetry of the single-ring \mbox{$e$-like} data 
in comparison with the best fit MC expectation and the expectations for $\mbox{sin}^{2}\theta_{23} = 0.4$
and $0.6$ appears in Fig~\ref{fig:sk1o2o3q23asymmetry}.


\subsection{$\theta_{13}$ analysis}

\begin{figure*}[ht]
  \begin{minipage}{6.5in}
    \includegraphics[height=2.5in, keepaspectratio=true, type=pdf,ext=.pdf,read=.pdf]{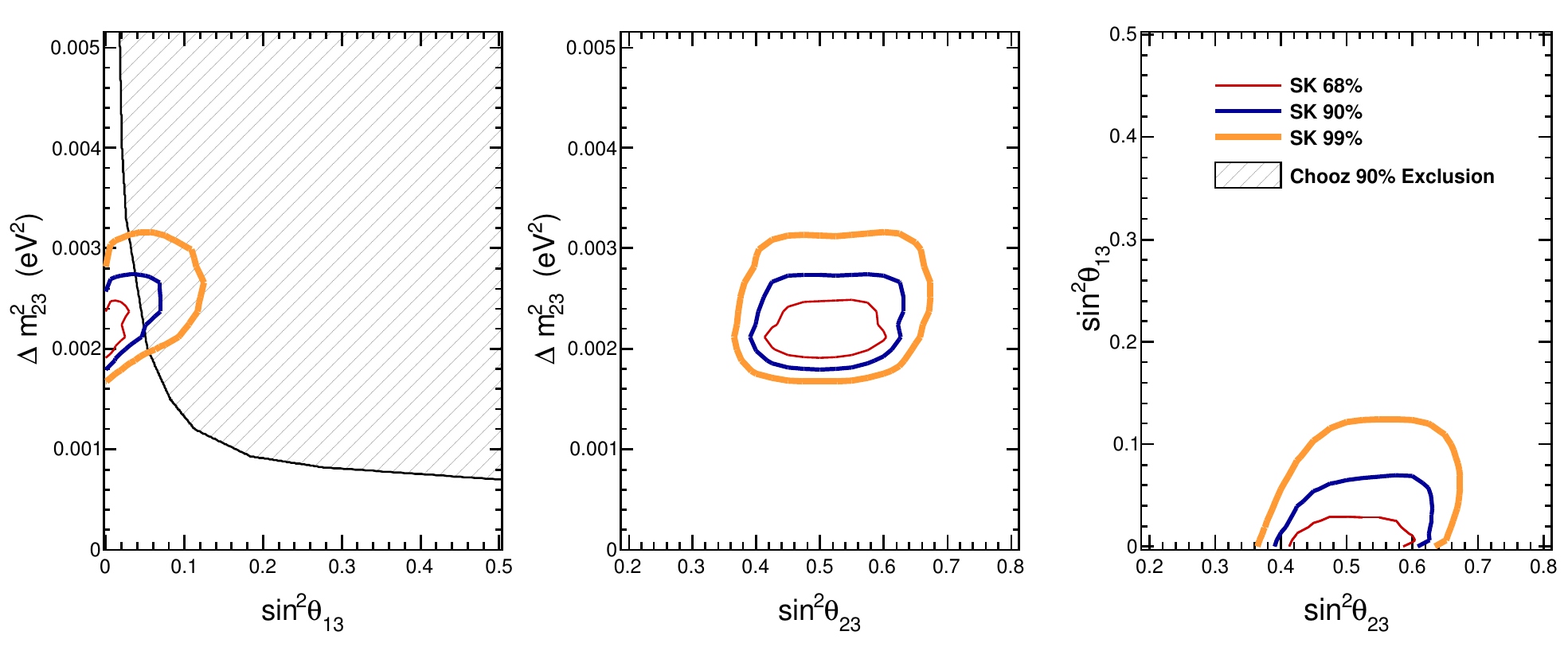}
    \caption{ 
        (color online).
              Normal hierarchy allowed regions at 68\% (thin line), 90\% (medium ), and 99\% (thick) C.L. 
              for the SK-I+II+III data.  
              The shaded region in the first panel shows 
              the Chooz 90\% exclusion region. 
     }
  \label{fig:normal.sk1o2o3}
  \end{minipage}
  \hfill
\end{figure*}

\begin{figure*}[ht]
  \begin{minipage}{6.5in}
    \includegraphics[height=2.5in, keepaspectratio=true, type=pdf,ext=.pdf,read=.pdf]{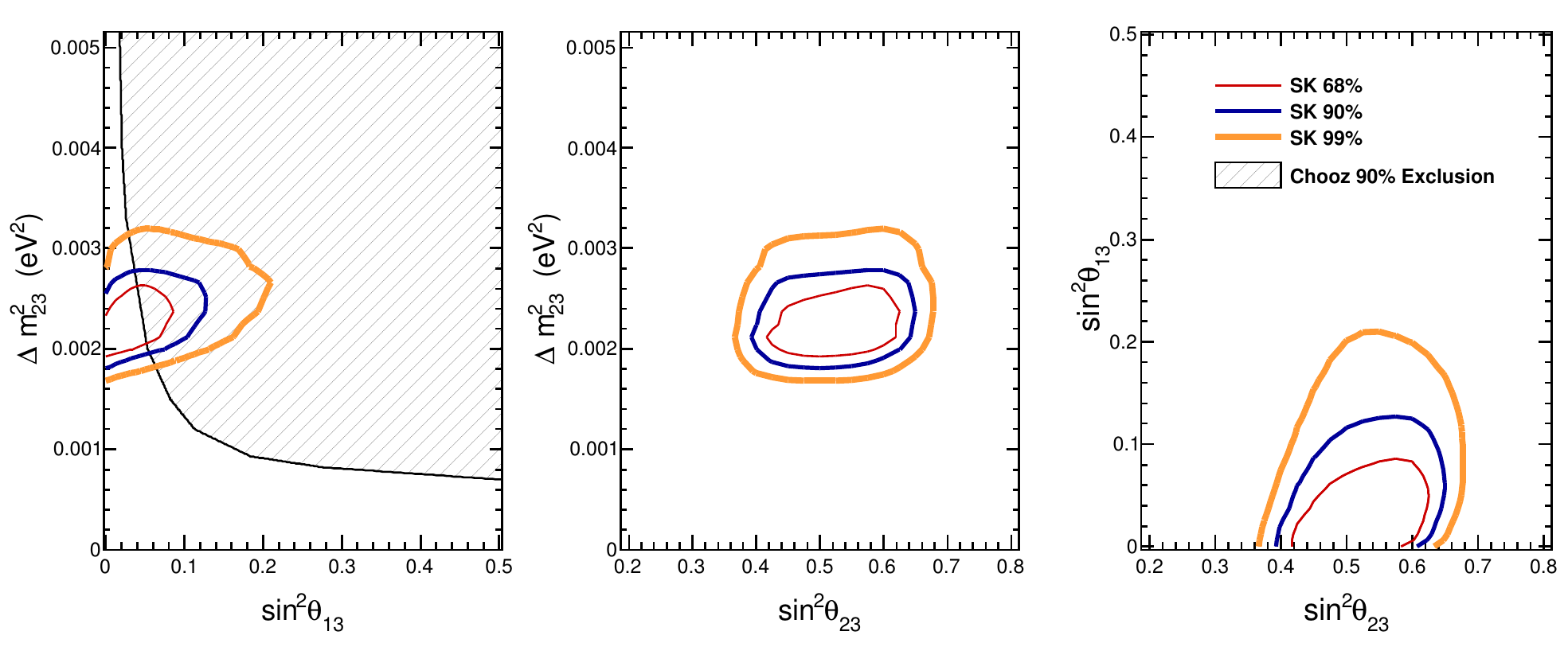}
    \caption{ 
        (color online).
              Inverted hierarchy allowed regions at 68\% (thin line), 90\% (medium), and 99\% (thick) C.L. 
              for the SK-I+II+III data.  
              The shaded region in the first panel shows 
              the Chooz 90\% exclusion region. 
            }
  \label{fig:inverted.sk1o2o3}
  \end{minipage}
  \hfill
\end{figure*}

\begin{figure*}[ht]
  \begin{minipage}{6.5in}
    \includegraphics[height=2.5in, keepaspectratio=true, type=pdf,ext=.pdf,read=.pdf]{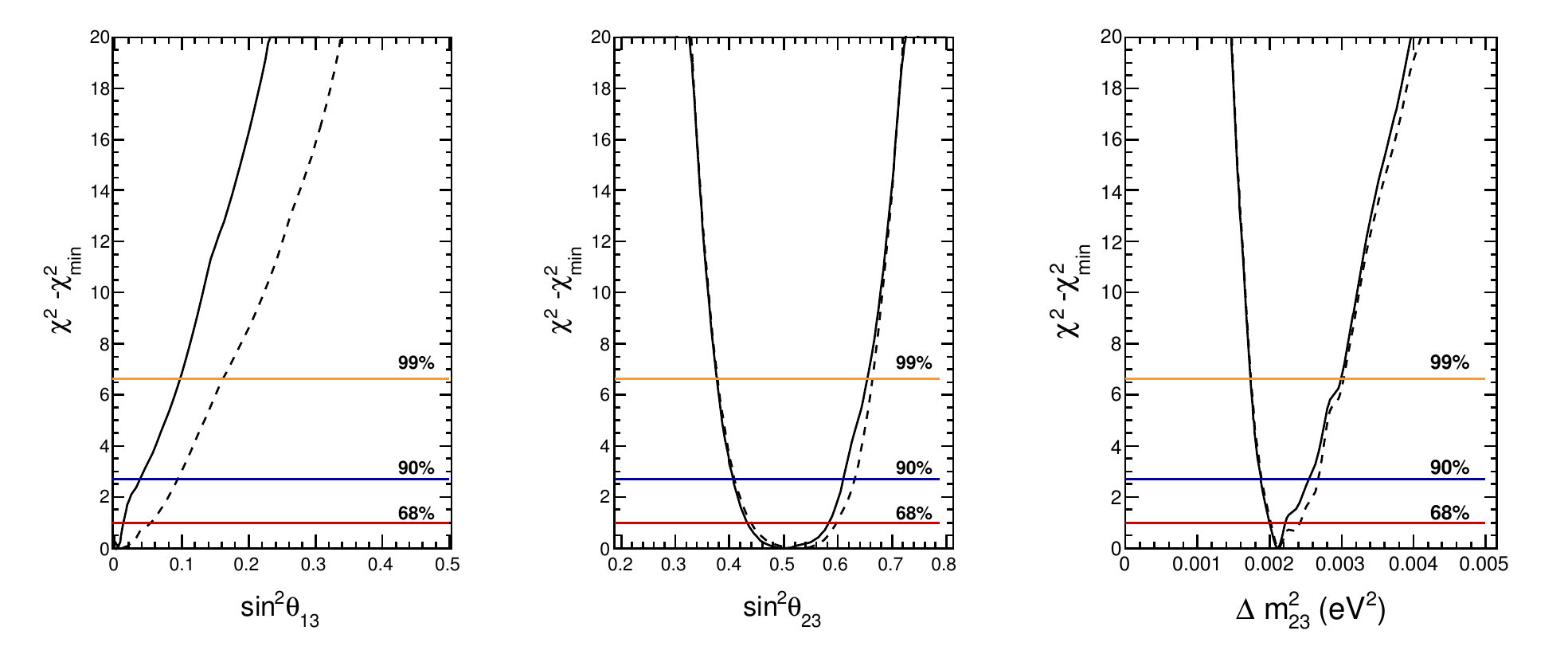}
    \caption{ 
        (color online).
              The $\Delta \chi^{2}$ distributions for the SK-I+II+III data set in the normal (solid) and 
              inverted (dashed) hierarchies. The horizontal lines represent the position of the one-dimensional
              cut value corresponding to the (from top to bottom) 99\%, 90\%, and 68\% CL. 
            }
  \label{fig:sk1o2o3.deltachi2}
  \end{minipage}
  \hfill
\end{figure*}

\begin{figure*}
  \begin{minipage}{6.5in}
    \includegraphics[width=5.5in,keepaspectratio=true,type=pdf,ext=.pdf,read=.pdf]{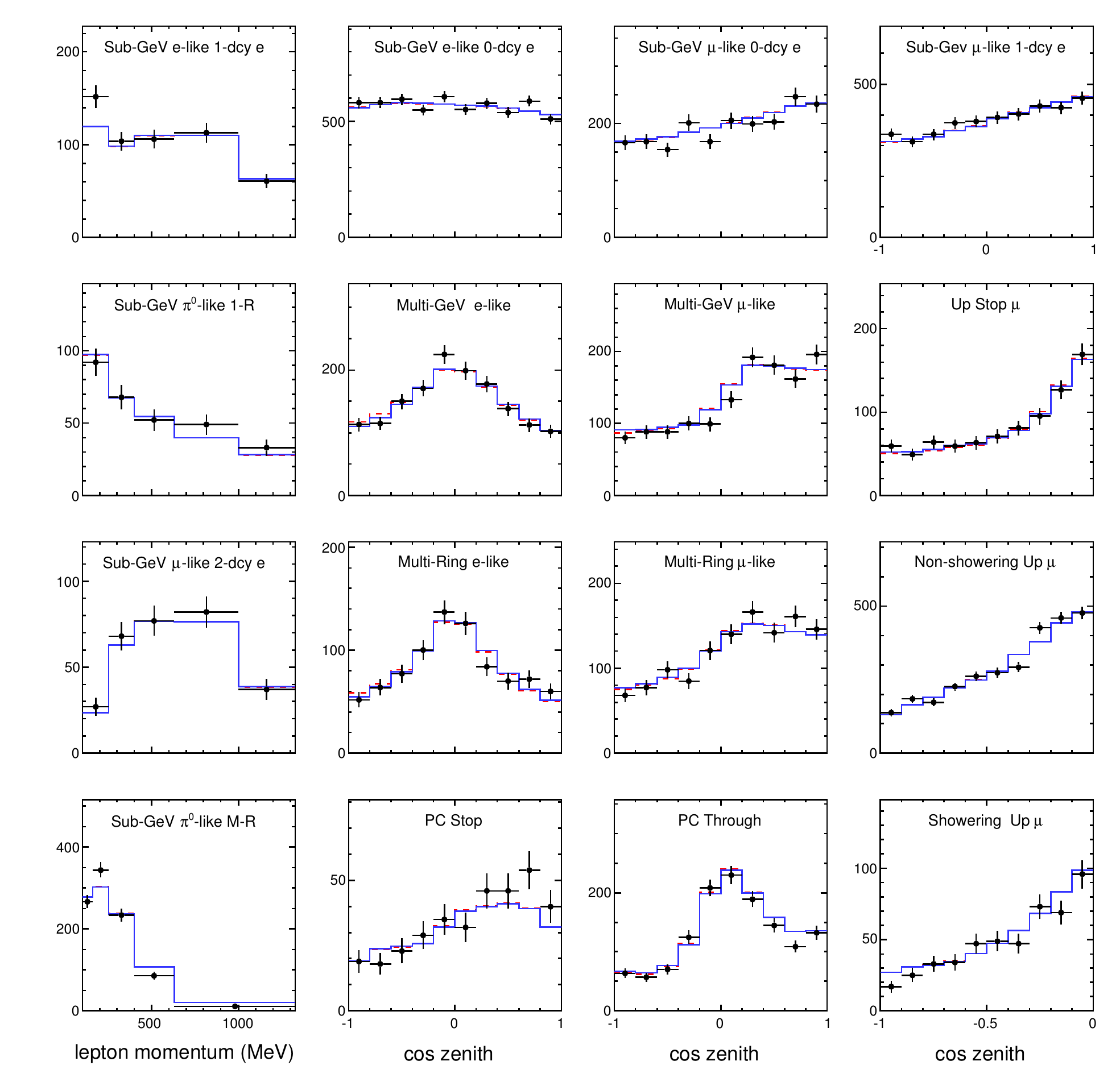}
  \end{minipage}
    \caption{ 
        (color online).
              SK-I+II+III zenith angle and lepton momentum distributions of the event samples used in the analyses. 
              Black dots represent the data with statistical errors,
              the solid lines are the MC expectation at the best fit from the $\theta_{13}$ analysis, and dashed lines 
              show the expectation at the best fit atmospheric variables but with $\theta_{13}$ at the Chooz limit. 
            }
  \label{fig:sk1o2o3samples}
\end{figure*}

\begin{figure*}[ht]
  \begin{minipage}{6.5in}
    \includegraphics[width=2.5in, height=2.5in,type=pdf,ext=.pdf,read=.pdf]{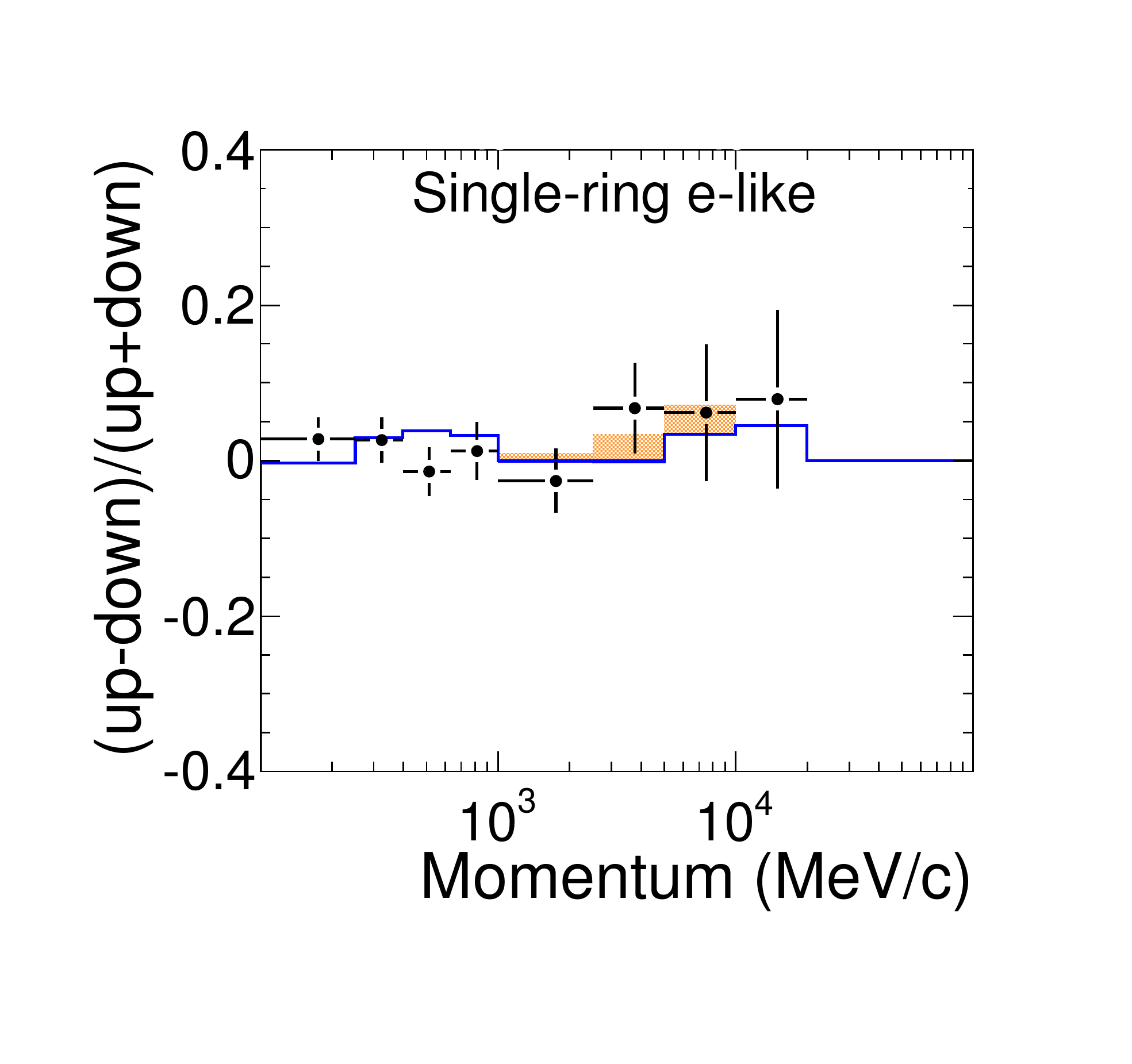}
    \includegraphics[width=2.5in, height=2.5in,type=pdf,ext=.pdf,read=.pdf]{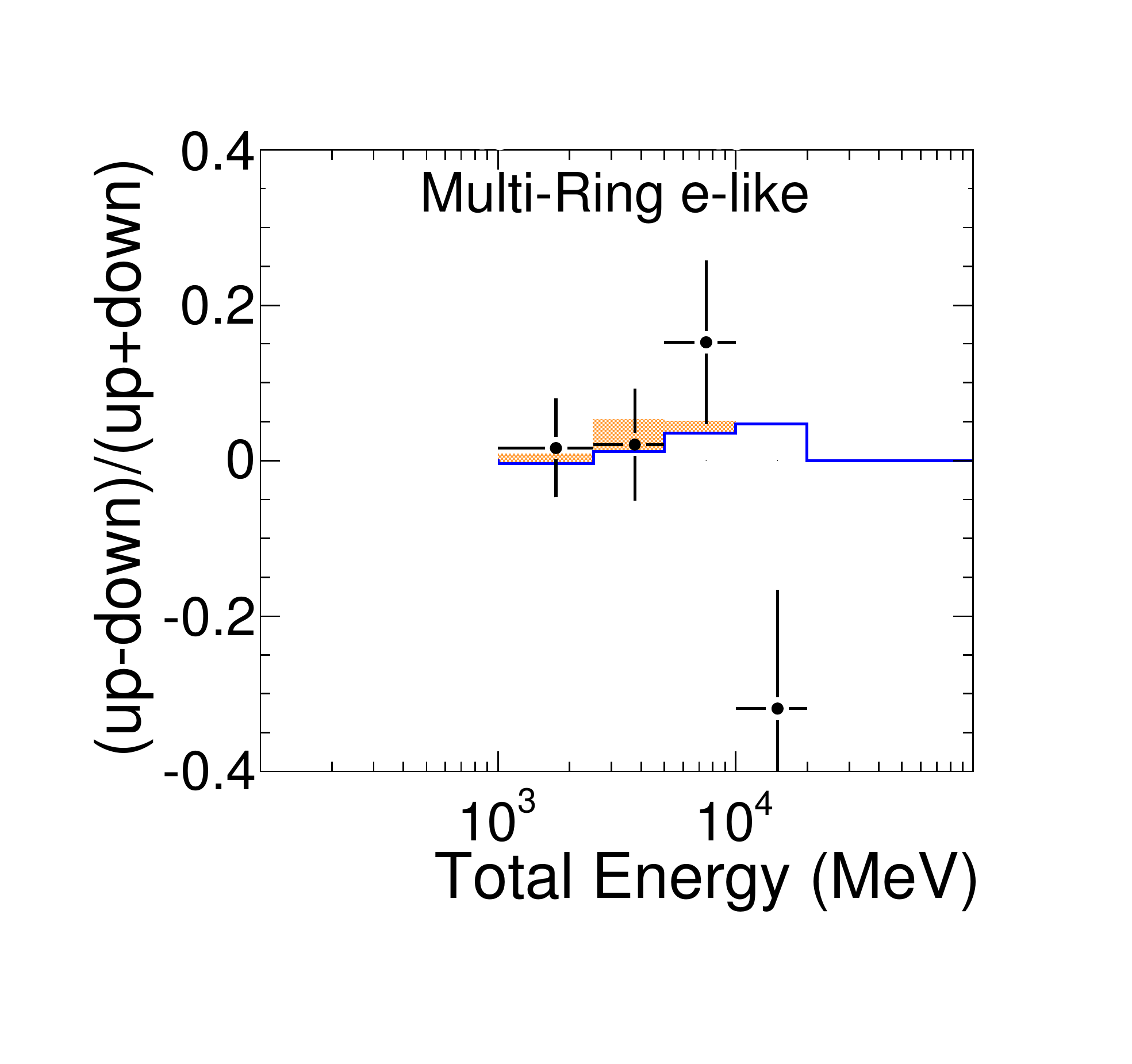}
  \end{minipage}
    \caption{ 
        (color online).
              Asymmetry (Up - Down)/(Up + Down) for the SK-I+II+III data set. Up is defined as events with 
              $\mbox{cos}\Theta < - 0.2$ and down as $\mbox{cos}\Theta > 0.2.$
              Solid lines represent the MC expectation at the best fit from the $\theta_{13}$ analysis and the shaded 
              regions show the expectation at the best fit atmospheric variables 
              but with $\theta_{13}$ at the Chooz limit for
              the single- (multi-) ring \mbox{$e$-like} events at left(right). Error bars are statistical.
            }
  \label{fig:sk1o2o3asymmetry}
\end{figure*}

  In the \tonethree analysis, oscillation fits are performed by scanning a grid 
of 83,025 oscillation points in three variables: 
$\mbox{log}_{10} \Delta m^{2}, \mbox{sin}^{2} \theta_{23}$, and $\mbox{sin}^{2} \theta_{13}$.
The fitting procedure has been performed on the combined SK-I+II+III data set assuming both 
a normal and inverted hierarchy.
The best fit in the normal hierarchy is at
$\Delta m^{2} = 2.1 \times 10^{-3} \mbox{eV}^{2}$, 
$\mbox{sin}^{2} \theta_{13} = 0.0$, and 
$\mbox{sin}^{2} \theta_{23} = 0.5$
with $\chi^{2}_{\mbox{min}} = 468.7.$ 
In the inverted hierarchy, the fit is at 
$\Delta m^{2} = 2.1 \times 10^{-3} \mbox{eV}^{2}$, 
$\mbox{sin}^{2} \theta_{13} = 0.006$, and  
$\mbox{sin}^{2} \theta_{23} = 0.53.$
The results are summarized in Tbl.~\ref{tbl:fits}. No preference is seen in the data for either mass hierarchy.
Confidence intervals at 90\%(99\%) are drawn in two dimensions at 
$\chi^{2} = \chi^{2}_{\mbox{min}} + 4.6 (9.2)$ and the third parameter point in these projections 
has been minimized over at each point in the plane to give the smallest value of $\chi^{2}$.
Computing the 90\% (99\%) critical value using a Feldman-Cousins~\cite{Feldman98} type procedure
confirmed that 4.6 (9.2) is the correct value in this scheme.
The resulting allowed regions and corresponding $\Delta \chi^{2}$ distributions are shown 
in Figs.~\ref{fig:normal.sk1o2o3}, ~\ref{fig:inverted.sk1o2o3}, and~\ref{fig:sk1o2o3.deltachi2}.
In the first panel of the former two figures, the $\Delta m^2$ vs. \tonethree plane for the fit
is overlaid with the Chooz~\cite{Chooz03} 90\% C.L. exclusion 
contour.
The zenith angle distributions of the combined data overlaid with the best fit MC expectation in the normal hierarchy and the expectation resulting from $\theta_{13}$ at the Chooz limit are shown in Fig.~\ref{fig:sk1o2o3samples}. Figure~\ref{fig:sk1o2o3asymmetry} shows the up-down asymmetry for the single-ring (multi-ring) \mbox{$e$-like} sample as a function of lepton momentum (total energy) for the single-ring (multi-ring) sample. The data are consistent with $\mbox{sin}^2 \theta_{13} = 0$.

\section{Conclusion}
\label{sec:conclusion}

   A three flavor oscillation fit to the first, second, and third generation Super-K atmospheric neutrino 
data has been performed. 
No evidence for $\theta_{13} > 0$ is found in fits to either hierarchy. 
The best fit oscillation parameters in the normal (inverted) hierarchy are $\Delta m^{2} = 2.1 \times 10^{-3}~\mbox{eV}^{2}$,
$\mbox{sin}^{2}\,\theta_{13} = 0.0 (0.006)$, and $\mbox{sin}^{2}\,\theta_{23} = 0.5 (0.53).$
The value of \tonethree is constrained to $\mbox{sin}^{2}\,\theta_{13} < 0.04 \xspace ( 0.09 )$ 
at the 90\% confidence level. All fits are consistent with the Chooz 
experiment's upper limit and no preference for either mass hierarchy exists in the data. 
The $\theta_{23}$ octant analysis finds no evidence for a preferred octant for \ttwothree.
However, the mixing angle is constrained at 90\% C.L. to $ 0.407 \le \mbox{sin}^{2} \theta_{23} \le 0.583.$

\section{Acknowledgments}

We gratefully acknowledge the cooperation of the Kamioka Mining and
Smelting Company.  The Super-Kamiokande experiment has been built and
operated from funding by the Japanese Ministry of Education, Culture,
Sports, Science and Technology, the United States Department of Energy,
and the U.S. National Science Foundation. Some of us have been supported by
funds from the Korean Research Foundation (BK21), and the Korea 
Science and Engineering Foundation. Some of us have been supported by 
the State Committee for Scientific Research in Poland (grant 1757/B/H03/2008/35).

\bibliography{bibliography}

\subsection{Appendix}
Tables~\ref{tbl:systcommonflux}, \ref{tbl:systcommonint}, and~\ref{tbl:systseparate} summarize the best fit systematic error parameters for the best fit point in the normal hierarchy fit from the $\theta_{13}$ analysis. 

%
\begin{table*}[htb]
   \begin{center}
   \begin{minipage}{\textwidth}
   \begin{tabular}{lllrr}
    \hline \hline
\multicolumn{3}{l}{   Systematic Error } &  \multicolumn{1}{c}{fit value} &  \multicolumn{1}{c}{ $\sigma$ }   \\
    \hline
Flux normalization    &     $E_{\nu} < 1$ GeV             &            &   34.7     &    25\footnote[1]
         {Uncertainty linearly decreases with $\log E_{\nu}$ from 25\,\%(0.1\,GeV) to 7\,\%(1\,GeV).} \\ 
                      &     $E_{\nu} > 1$ GeV             &            &    8.8     &     7\footnote[2]
         {Uncertainty is 7\,\% up to 10\,GeV, linearly increases with $\log E_{\nu}$ from 7\,\%(10\,GeV) to 12\,\%(100\,GeV) and then to 20\,\%(1\,TeV)} \\
  $\nu_{\mu}/\nu_{e}$ &                                   &            &            &       \\
                      &     $E_{\nu} < 1$ GeV             &            &   -1.9     &     2 \\
                      &     $1 < E_{\nu} < 10$ GeV        &            &   -2.5     &     3 \\
                      &     $E_{\nu} > 10$ GeV            &            &   -3.7     &     5\footnote[3]
         {Uncertainty linearly increases with $\log E_{\nu}$ from 5\,\%(30\,GeV) to 30\,\%(1\,TeV).}\\
 $\bar \nu_{e}/\nu_{e}$ &                                 &            &            &       \\
                      &     $E_{\nu} < 1 $ GeV            &            &    5.54    &    5  \\
                      &     $1 < E_{\nu} < 10 $ GeV       &            &    1.13    &    5  \\
                      &     $E_{\nu} > 10$ GeV            &            &   -0.10    &    8\footnote[4]
         {Uncertainty linearly increases with $\log E_{\nu}$ from 8\,\%(100\,GeV) to 20\,\%(1\,TeV).}\\
$\bar \nu_{\mu}/\nu_{\mu}$ &                              &            &            &       \\
                      &     $E_{\nu} < 1$ GeV             &            &   -0.48    &    2  \\ 
                      &     $1 < E_{\nu} < 10 $ GeV       &            &   -1.35    &    6  \\
                      &     $E_{\nu} > 10$ GeV            &            &   -1.75    &    6\footnote[5]
         {Uncertainty linearly increases with $\log E_{\nu}$ from 6\,\%(50\,GeV) to 40\,\%(1\,TeV).}\\ 
Up/down ratio         & $<$ 400 MeV          &  $e$-like               &    -0.07   &        0.1  \\
                      &                      &  $\mu$-like             &    -0.23   &        0.3  \\   
                      &                      &  0-decay $\mu$-like     &    -0.84   &        1.1  \\
                      & $>$ 400 MeV          &  $e$-like               &    -0.61   &        0.8  \\
                      &                      &  $\mu$-like             &    -0.38   &        0.5  \\
                      &                      &  0-decay $\mu$-like     &    -1.29   &        1.7  \\
                      & Multi-GeV            &  $e$-like               &    -0.53   &        0.7  \\
                      &                      &  $\mu$-like             &    -0.15   &        0.2 \\
                      & Multi-ring Sub-GeV   &  $\mu$-like             &    -0.15   &        0.2 \\
                      & Multi-ring Multi-GeV &  $e$-like               &    -0.23   &        0.3 \\
                      &                      &  $\mu$-like             &    -0.15   &        0.2 \\
                      & PC                   &                         &    -0.15   &        0.2 \\
Horizontal/Vertical ratio & $<$ 400 MeV      &  $e$-like               &    -0.01   &        0.1  \\
                      &                      &  $\mu$-like             &    -0.01   &        0.1  \\   
                      &                      &  0-decay $\mu$-like     &    -0.03   &        0.3  \\
                      & $>$ 400 MeV          &  $e$-like               &    -0.14   &        1.4  \\
                      &                      &  $\mu$-like             &    -0.19   &        1.9  \\
                      &                      &  0-decay $\mu$-like     &    -0.14   &        1.4  \\
                      & Multi-GeV            &  $e$-like               &    -0.33   &        3.2 \\
                      &                      &  $\mu$-like             &    -0.23   &        2.3 \\
                      & Multi-ring Sub-GeV   &  $\mu$-like             &    -0.13   &        1.3 \\
                      & Multi-ring Multi-GeV &  $e$-like               &    -0.29   &        2.8 \\
                      &                      &  $\mu$-like             &    -0.15   &        1.5 \\
                      & PC                   &                         &    -0.17   &        1.7 \\
\multicolumn{3}{l}{  K/$\pi$ ratio in flux calculation     }           &   -12.9    &   10\footnote[6]
	    {Uncertainty increases linearly from 5$\%$ to 20$\%$ between 100GeV and 1TeV.}\\
\multicolumn{3}{l}{  Neutrino path length                          } &    -8.8    &        10  \\
Sample-by-sample      & FC Multi-GeV         &                       &    -4.5    &         5  \\
                      & PC + Up-stop $\mu$   &                       &    -7.1    &         5  \\
  \hline
  \hline
  \end{tabular}
  \end{minipage}
 \caption{Flux-related systematic errors that are common to all SK geometries. 
          The second column shows the best fit value
          of the systematic error parameter $\epsilon_{i}$ in percent and the third column shows
          the estimated 1-$\sigma$ error size in percent.}
 \label{tbl:systcommonflux}
\end{center}
\end{table*}

%
\begin{table*}[htb]
   \begin{center}
   \begin{minipage}{\textwidth}
   \begin{tabular}{lllrr}
    \hline \hline
\multicolumn{3}{l}{   Systematic Error } &  \multicolumn{1}{c}{fit value} &  \multicolumn{1}{c}{ $\sigma$ }   \\
    \hline

\multicolumn{3}{l}{  MA in QE and single $\pi$                     } &   -2.4     &        10 \\
\multicolumn{3}{l}{  CCQE cross section                            } &    0.66    &  1.0\footnote[1]
	    {Difference from the Nieves~\cite{Nieves04} model is set to 1.0}\\
\multicolumn{3}{l}{  Single meson production cross section         } &    7.8     &        20  \\
\multicolumn{3}{l}{  DIS cross section ($E_{nu} <$ 10 GeV)         } &   -0.16    &  1.0\footnote[2]
	    {Difference from CKMT~\cite{CKMT94} parametrization is set to 1.0}\\
\multicolumn{3}{l}{  DIS cross section                             } &   2.27     &         5  \\
\multicolumn{3}{l}{  Coherent $\pi$ production                     } &   1.53     &        100 \\
\multicolumn{3}{l}{  NC/(CC)                                       } &   1.51     &         20 \\ 
\multicolumn{3}{l}{  Nuclear effect in $^{16}$O nucleus            } &   -13.8    &         30 \\
\multicolumn{3}{l}{  Nuclear effect in pion spectrum               } &    0.8     &   1.0\footnote[3]
	    {Difference between NEUT~\cite{neut} and NUANCE~\cite{nuance} is set to 1.0}\\
\multicolumn{3}{l}{  $\nu_{\tau}$ contamination                    } &    1.0     &         30 \\
\multicolumn{3}{l}{  NC in FC $\mu$-like (hadron simulation)       } &   -4.6     &         10 \\
\multicolumn{3}{l}{  CCQE $\bar \nu_{i} /\nu_{i}$ (i=e,$\mu$) ratio} &   0.84     &    1.0\footnotemark[1] \\
\multicolumn{3}{l}{  CCQE $\mu$/e ratio                            } &   1.12     &    1.0\footnotemark[1] \\
\multicolumn{3}{l}{  Single $\pi$ production, $\pi^{0}/\pi^{\pm}$ ratio  }        &   -29.0  &  40 \\
\multicolumn{3}{l}{  Single $\pi$ production, $\bar \nu_{i} / \nu_{i}$ (i=e,$\mu$) ratio} & -0.04 &  1.0\footnote[4]
	    {Difference from the Hernandez\cite{Hernandez07} model is set to 1.0}\\
$\pi^{+}$ decay uncertainty        
                     & Sub-GeV 1-ring & $e$-like   0-decay           &  -0.48     &     0.5  \\ 
                     &         & $\mu$-like   1-decay                &   0.77     &    -0.8  \\
                     &         & $e$-like   1-decay                  &   3.9      &    -4.1  \\
                     &         & $\mu$-like   0-decay                &  -0.77     &     0.8  \\
                     &         & $\mu$-like   2-decay                &   5.46     &    -5.7  \\
  \hline
  \hline
  \end{tabular}
  \end{minipage}
 \caption{Neutrino interaction and particle production systematic errors that are common to all SK geometries. 
          The second column shows the best fit value
          of the systematic error parameter $\epsilon_{i}$ in percent and the third column shows
          the estimated 1-$\sigma$ error size in percent.}
 \label{tbl:systcommonint}

\end{center}
\end{table*} 

\begin{table*}[htb]
   \begin{center}
   \begin{minipage}{\textwidth}
   \begin{tabular}{lllrrrrrr}
   \hline \hline
   \multicolumn{3}{c}{}  & \multicolumn{2}{c}{SK-I}  & \multicolumn{2}{c}{SK-II}  & \multicolumn{2}{c}{SK-III}  \\
   \multicolumn{3}{l}{Systematic Error}    & fit   & $\sigma$& fit    & $\sigma $  & fit    & $\sigma $  \\
   \hline
\multicolumn{3}{l}{FC reduction            }                 &    0.005 & 0.2  &    0.008 & 0.2  &  0.061  &  0.8   \\
\multicolumn{3}{l}{PC reduction            }                 &   -0.99  & 2.4  &   -2.12  & 4.8  &  0.034  &  0.5   \\
\multicolumn{3}{l}{FC/PC separation            }             &   -0.058 & 0.6  &   0.068  & 0.5  & -0.28   &  0.9   \\ 
\multicolumn{3}{l}{PC-stop/PC-through separation (top)   }   &    7.84  & 14   &  -17.47  &21    &-20.03   & 31     \\ 
\multicolumn{3}{l}{PC-stop/PC-through separation (barrel)}   &   -2.27  & 7.5  &  -31.51  &17    &  3.44   & 23     \\ 
\multicolumn{3}{l}{PC-stop/PC-through separation (bottom)}   &   -2.32  & 11.  &   -7.32  &12    &  1.59   & 11     \\ 
Non-$\nu$ BG ($e$-like)                                  
                      &  Sub-GeV             &               &  0.077   &  0.5 &   0.004  &  0.2 &   0.003 &  0.1 \\ 
                      &  Multi-GeV           &               &  0.047   &  0.3 &   0.005  &  0.3 &   0.011 &  0.4 \\
Non-$\nu$ BG ($\mu$-like)                                
                      &  Sub-GeV             &               & -0.01    &  0.1 &   0.02   &  0.1 &   0.052 &  0.1 \\ 
                      &  Multi-GeV           &               & -0.01    &  0.1 &   0.02   &  0.1 &   0.11  &  0.2 \\      
                      &  Sub-GeV 1-ring & $\mu$-like 0-decay & -0.04    &  0.4 &   0.02   &  0.1 &   0.052 &  0.1 \\ 
                      &  PC                  &               & -0.02    &  0.2 &   0.14   &  0.7 &   0.95  &  1.8 \\
\multicolumn{3}{l}{Fiducial volume                       }   &   -0.23  & 2    &   0.43   & 2    &   0.93  &  2    \\ 
Ring separation                                           
                      & $<$ 400 MeV          &  $e$-like     &  1.23    &  2.3 &  -1.67   &  1.3 &   0.12   &  2.3  \\ 
                      &                      &  $\mu$-like   &  0.37    &  0.7 &  -2.96   &  2.3 &   0.037  &  0.7  \\
                      & $>$ 400 MeV          &  $e$-like     &  0.21    &  0.4 &  -2.19   &  1.7 &   0.021  &  0.4  \\
                      &                      &  $\mu$-like   &  0.37    &  0.7 &  -0.90   &  0.7 &   0.036  &  0.7  \\
                      & Multi-GeV            &  $e$-like     &  1.97    &  3.7 &  -3.35   &  2.6 &   0.19   &  3.7  \\
                      &                      &  $\mu$-like   &  0.91    &  1.7 &  -2.19   &  1.7 &   0.089  &  1.7  \\
                      & Multi-ring sub-GeV   &  $\mu$-like   & -2.40    & -4.5 &  10.56   & -8.2 &  -0.24   & -4.5  \\
                      & Multi-ring multi-GeV &  $e$-like     &  0.05    &  0.1 &  -2.45   &  1.9 &   0.16   &  3.1   \\
                      &                      &  $\mu$-like   & -2.19    & -4.1 &   1.03   & -0.8 &  -0.21   & -4.1  \\
Particle identification                   
                      & Sub-GeV              &  $e$-like     & -0.007   &  0.1 &   0.13   &  0.5 &   0.004  &  0.1  \\
                      &                      &  $\mu$-like   &  0.007   & -0.1 &  -0.13   & -0.5 &  -0.004  & -0.1  \\
                      & Multi-GeV            &  $e$-like     & -0.014   &  0.2 &   0.023  &  0.1 &   0.008  &  0.2  \\
                      &                      &  $\mu$-like   &  0.014   & -0.2 &  -0.023  & -0.1 &  -0.008  & -0.2  \\
Particle identification (multi-ring)                       
                      & Sub-GeV              &  $\mu$-like   & -0.18    & -3.9 &  -0.55   & -2.2 &  -0.15   & -3.9 \\
                      & Multi-GeV            &  $e$-like     &  0.078   &  1.7 &   0.45   &  1.8 &   0.063  &  1.7 \\
                      &                      &  $\mu$-like   & -0.13    & -2.9 &  -0.86   & -3.4 &  -0.11   & -2.9 \\
\multicolumn{3}{l}{Energy calibration                     }  &  -0.002  & 1.1  &   -0.56  & 1.7  &  -0.35  &  2.7  \\ 
\multicolumn{3}{l}{Up/Down asymmetry energy calibration   }  &   -0.4   & 0.6  &   -0.15  & 0.6  &  -0.03  &  1.3  \\ 
Upward-going muon reduction                              
                      &  Stopping            &               & -0.057   & 0.7  &  -0.14   &  0.7 &   0.14  &  0.7 \\
                      &  Through-going       &               & -0.041   & 0.5  &  -0.10   &  0.5 &   0.10  &  0.5 \\
\multicolumn{3}{l}{Upward stopping/through-going $\mu$ separation  }
                                                             &   -0.04  & 0.4  &    0.006 & 0.4  &   0.04  &  0.6  \\ 
\multicolumn{3}{l}{Energy cut for upward stopping $\mu$   }         & -0.13   &  0.8 &  -0.26   &  1.4 & 0.78 &  2.1  \\ 
\multicolumn{3}{l}{Path length cut for upward through-going $\mu$}  &  0.39   &  1.8 &  -1.0    &  2.1 & 0.4  &  1.6  \\ 
\multicolumn{3}{l}{Upward through-going $\mu$ showering separation} &  9.42   &  9.0 &   2.28   & 13.0 & 6.1  &  6.0  \\ 
BG subtraction of upward $\mu$ \footnote[1]
{The uncertainties in BG subtraction for upward-going muons are only for the most horizontal bins  $-0.1 < \cos\theta < 0$.}
                      &\multicolumn{2}{l}{Stopping}          &  4.16    & 16   &  -7.47   & 21   &   0.004 & 20   \\ 
                      &\multicolumn{2}{l}{Non-showering}     & -1.24    & 11   &   8.08   & 15   &   6.34  & 19   \\ 
                      &\multicolumn{2}{l}{Showering}         &  2.27    & 18   & -18.16   & 14   &  24.7   & 24   \\ 
\multicolumn{3}{l}{Multi-GeV Single-Ring Electron BG      }  &  5.95    & 16.3 &  -4.67   & 23.4 &   1.06  & 41.4\\ 
\multicolumn{3}{l}{Multi-GeV Multi-Ring Electron BG       }  & -4.38    & 35.6 &  -1.4    & 22.3 & -16.8   & 38.0\\ 
\multicolumn{3}{l}{Multi-GeV Multi-Ring $e$-like likelihood }& -1.12    &  6.4 &   0.5    & 11.1 &  -0.3   &  5.3\\ 
Sub-GeV 1-ring $\pi^{0}$ selection                         
                     &$\;\,100 < \mbox{P}_{e}<\;\,250$ &MeV/c& -3.94    & 11.2 &  -4.08   &  7.5 &  -5.34  &  7.7  \\ 
                     &$\;\,250 < \mbox{P}_{e}<\;\,400$ &     & -4.05    & 11.5 &  -4.85   &  8.9 & -18.37  & 26.4  \\
                     &$\;\,400 < \mbox{P}_{e}<\;\,630$ &     & -8.23    & 23.4 &  -9.52   & 17.5 &  -8.70  & 12.5  \\
                     &$\;\,630 < \mbox{P}_{e}<1000$ &        & -6.72    & 19.1 &  -5.81   & 10.7 & -18.58  & 26.7  \\
                     &$1000 < \mbox{P}_{e} <1330$ &          & -4.57    & 13.0 &  -6.03   & 11.1 & -18.58  & 26.7  \\
\multicolumn{3}{l}{Sub-GeV 2-ring $\pi^{0}$              }   &   -0.31  & 2    &   0.024  & 2    &   0.009 &  1    \\ 
Decay-e tagging                                            
                     &                &                      &   0.16   & 1.5  &   0.41   & 1.5  &   1.06  & 1.5   \\
\multicolumn{3}{l}{Solar Activity                        }   &   0.6    & 20   &   27.9   & 50   &   3.78  &  20   \\ 
  \hline
  \hline
  \end{tabular}
  \end{minipage}
 \caption{Systematic errors that are independent between SK-I, SK-II, and SK-III. Columns labeled ``fit'' show the best
          fit value of the systematic error parameter $\epsilon_{i}$ in percent. Those labeled ``$\sigma$''
          show the estimated 1-$\sigma$ error size in percent.
         }
 \label{tbl:systseparate}
\end{center}
\end{table*}

\end{document}